\documentclass[aps,prx,article,twocolumn,preprintnumbers,amsmath,amssymb,superscriptaddress]{revtex4-1}

\date{\today}
\usepackage{epsfig}
\usepackage{subfigure}
\usepackage{graphicx}
\usepackage{dcolumn}
\usepackage{bm}
\usepackage[colorlinks,linkcolor=blue,hyperindex,CJKbookmarks]{hyperref}
\usepackage{float}
\usepackage{hyperref}
\usepackage{comment}
\usepackage{color}
\newcommand{\tJ}{$t\small{-}J$}

\newcommand{\Akw}{$A({\bf k},\omega)$}

\begin{document}
\title{Emergence of Quasiparticles in a Doped Mott Insulator}
\author{Yao Wang}
\email{yaowang@g.clemson.edu}
\affiliation{Department of Physics and Astronomy, Clemson University, Clemson, South Carolina 29631, USA}
 \author{Yu He}
 \affiliation{Department of Physics, University of California at Berkeley, Berkeley, California 94720, USA}
\affiliation{Stanford Institute for Materials and Energy Sciences, SLAC National Laboratory and Stanford University, Menlo Park, CA 94025, USA}
\author{Krzysztof Wohlfeld}
\affiliation{Institute of Theoretical Physics, Faculty of Physics, University of Warsaw, Pasteura 5, PL-02093 Warsaw, Poland}
\author{Makoto Hashimoto}
\affiliation{Stanford Synchrotron Radiation Lightsource, SLAC National Accelerator Laboratory, Menlo Park, CA 94025, USA}
\author{Edwin W. Huang}
 \affiliation{Department of Physics and Department of Applied Physics, Stanford University, Stanford, CA 94305, USA}
 \affiliation{Stanford Institute for Materials and Energy Sciences, SLAC National Laboratory and Stanford University, Menlo Park, CA 94025, USA}
\affiliation{Department of Physics and Institute of Condensed Matter Theory, University of Illinois at Urbana-Champaign, Urbana, 61801 Illinois, USA}
\author{Donghui Lu}
\affiliation{Stanford Synchrotron Radiation Lightsource, SLAC National Accelerator Laboratory, Menlo Park, CA 94025, USA}
\author{Sung-Kwan Mo}
\affiliation{Advanced Light Source, Lawrence Berkeley National Laboratory, Berkeley, CA 94720, USA}
\author{Seiki Komiya}
\affiliation{Central Research Institute of Electric Power Industry, Yokosuka, Kanagawa 240-0196, Japan}
\author{Chunjing Jia}
\affiliation{Stanford Institute for Materials and Energy Sciences, SLAC National Laboratory and Stanford University, Menlo Park, CA 94025, USA}
\author{Brian Moritz}
\affiliation{Stanford Institute for Materials and Energy Sciences, SLAC National Laboratory and Stanford University, Menlo Park, CA 94025, USA}
\author{Zhi-Xun Shen}
 \affiliation{Department of Physics and Department of Applied Physics, Stanford University, Stanford, CA 94305, USA}
\affiliation{Stanford Institute for Materials and Energy Sciences, SLAC National Laboratory and Stanford University, Menlo Park, CA 94025, USA}
\author{Thomas P. Devereaux}
\email{tpd@stanford.edu}
\affiliation{Stanford Institute for Materials and Energy Sciences, SLAC National Laboratory and Stanford University, Menlo Park, CA 94025, USA}
\affiliation{Department of Materials Science and Engineering, Stanford University, Stanford, CA 94305 USA}
\date{\today}
\begin{abstract}
{\bf Abstract:} How a Mott insulator develops into a weakly coupled metal upon doping is a central question to understanding various emergent correlated phenomena. To analyze this evolution and its connection to the high-$T_c$ cuprates, we study the single-particle spectrum for the doped Hubbard model using cluster perturbation theory on superclusters. Starting from extremely low doping, we identify a heavily renormalized quasiparticle dispersion that immediately develops across the Fermi level, and a weakening polaronic side band at higher binding energy.  The quasiparticle spectral weight roughly grows at twice the rate of doping in the low doping regime, but this rate is halved at optimal doping. In the heavily doped regime, we find both strong electron-hole asymmetry and a persistent presence of Mott spectral features. Finally, we discuss the applicability of the single-band Hubbard model to describe the evolution of nodal spectra measured by angle-resolved photoemission spectroscopy (ARPES) on the single-layer cuprate La$_{2-x}$Sr$_x$CuO$_4$ ($0 \le x \le 0.15$). This work benchmarks the predictive power of the Hubbard model for electronic properties of high-$T_c$ cuprates. 
\end{abstract}
\maketitle

\section{Introduction}
One of the most important questions in the field of quantum materials is the nature of states that emerge from a doped Mott insulator\,\cite{Wen1996, lee2006doping}. This is a key prerequisite to understanding the development of high-$T_c$ superconductivity via hole or electron doping from the Mott phase\,\cite{Stephan1991,Dagotto:1994cz}. Furthermore, this helps elucidate why and how the cuprates are different from other doped Mott insulators. With ability to resolve single-particle energy-momentum spectra, angle-resolved photoemission (ARPES) has helped to determine a starting point to modelling cuprates -- the doped Hubbard model\,\cite{Damascelli2003}. Early on it was shown that strong correlations, in the form of the on-site Hubbard $U$, lead to an antiferromagnetic (AFM) state from a predominantly Cu $3d^9$ configuration\,\cite{zhang1988effective}. When a single hole is created by photoemission, it disperses from ($\pi$/2, $\pi$/2) towards the $\Gamma$ point, which abruptly falls in intensity near the zone center -- the so-called ``waterfall''\,\cite{ronning2005anomalous, meevasana2007hierarchy, valla2007high, xie2007high}. Rather than dispersing as a free quasiparticle, the low-energy band in a Mott insulator can be well fitted with a velocity on the scale of the spin exchange $J\sim$120\,meV\,\cite{Martinez:1991SpinPolaron, Bala:1995ttprJSCBA, macridin2007high, Efstratios:2007SCBA}. These basic observations of a Mott insulator lie within the framework of the Hubbard model.

In doped Mott insulators, recent experimental and theoretical progress further extends Hubbard model's descriptiveness. On the experimental side, ARPES and resonant inelastic x-ray scattering (RIXS) have shown evidence of strong correlations\,\cite{graf2007universal, he2019fermi,he2018rapid, chen2019strange} and collective spin excitations\,\cite{le2011intense, dean2013persistence, dean2013high, lee2014asymmetry, ishii2014high} far beyond the Mott phase, which cannot be addressed using weak-coupling theory. 

Despite such success, clearly, there are also experimental observations that seem to lie beyond the simple Hubbard and \tJ\ models. An outstanding example is the $\sim200$\,meV linewidth of the ARPES spectrum even at ($\pi$/2, $\pi$/2)\,\cite{ronning2005anomalous,shen2005nodal}, where a single hole described by the \tJ\ model has no phase space to decay\,\cite{Martinez:1991SpinPolaron,Bala:1995ttprJSCBA}. The correct linewidth was obtained only by considering the lattice polaronic effect\,\cite{shen2004missing, mishchenko2004electron, mishchenko2006numerical, shen2007angle}, which becomes less dressed at higher dopings due to screening\cite{ rosch2005polaronic, slezak2006spectral, gunnarsson2006electron}. Moreover, in many doped Mott insulators, such as the nickelates\,\cite{tranquada1994simultaneous}, manganites\,\cite{ramirez1996thermodynamic}, and cobaltates\,\cite{cwik2009magnetic}, doped carriers cause insulating charge structures rather than superconductivity\,\cite{ulbrich2012neutron}. While stripe order has also been observed in cuprates, the magnitude is not nearly as strong as in other transition metal oxides\,\cite{tranquada2013spins}. This difference also led to the consideration of many material-specific degrees of freedom beyond the prototypical Mott insulator, including static and dynamic lattice effects\,\cite{bogdanov2000evidence, lanzara2001evidence, zhou2003high, cuk2004coupling, johnston2012evidence, ohta1991apex, peng2017influence, slezak2008imaging, sakakibara2012origin}. Therefore, both for aspects specific to cuprates and those generic to correlated materials, one may wonder, to what extent exactly can the impact of correlations be captured within such toy models?\,\cite{kordyuk2005bare, yoshida2006systematic,  fournier2010loss}.

To identify universal features resulted from purely electron-electron correlations and their evolution with doping, we systematically study the single-particle spectral function of the doped Hubbard model, with only $t$, $t^\prime$ and $U$ and no other external ingredients. By calculating the spectral function at extremely fine doping levels over a wide range of electron and hole dopings, we expect to unbiasedly decipher the doping evolution of the dispersions, weights, and lineshapes of spectral features, and compare with ARPES experiments in cuprates. Though some of these properties have been studied previously at some discrete doping levels\cite{Kohno:2012PRL,PhysRevB.90.035111}, we find that the emergence of quasiparticles and their interplay with the polaronic Mott features cannot be fully addressed without reaching extreme limits of dopings. A detailed understanding of these spectral properties may provide insight on what aspects of experimental photoemission data can or cannot be well represented by the Hubbard model.

\section{Results and Discussion}

The Hubbard model is presented in the Method section. Historically, a variety of numerical techniques have been used to investigate the single-particle spectrum of the Hubbard model, \textit{e.g.}~exact diagonalization\,\cite{PhysRevB.44.10256, Dagotto1992, RevModPhys.70.1039}, quantum Monte Carlo\,\cite{white1989numerical,foulkes2001quantum}, density-matrix renormalization group\,\cite{white1992density,schollwock2005density}, dynamical mean-field theory\,\cite{kancharla2008anomalous, weber2010strength, gull2013superconductivity, sakai2009evolution}, CPT\,\cite{gros1993cluster,Senechal:2000fg, Senechal:2002fr, Kohno:2012PRL, PhysRevB.90.035111} and others\,\cite{yuan2005doping,arrigoni2009phase, han2016charge, hettler2000dynamical,potthoff2003variational}. To investigate low-temperature spectral features with fine momentum resolution and continuous doping dependence, CPT with superclusters is the most suitable approach. 

\begin{figure}
\begin{center}
\includegraphics[width=\columnwidth]{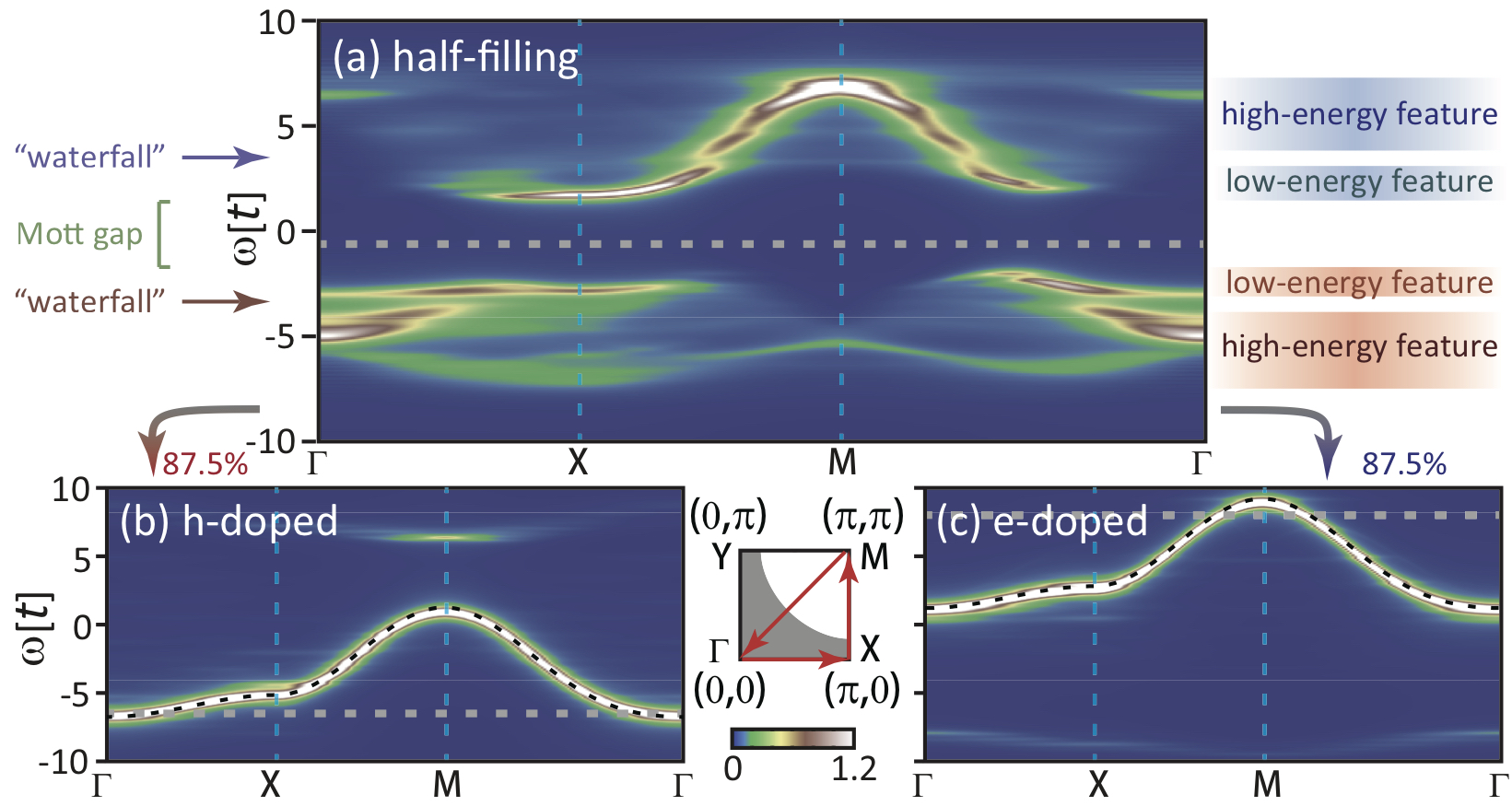}
\caption{\label{fig:1} {\bf Overview of spectral features. }
Spectral function of the Hubbard model calculated using cluster perturbation theory (CPT): (a) for the half-filled Mott insulator and (b,c) for the 87.5\% hole- and electron-doped system, respectively. The dotted line in (b) and (c) denotes the non-interacting tight-binding dispersion, while the horizontal dashed lines mark $E_{\rm F}$. The Brillouin zone (BZ) cartoon shows the spectral cut between the $\Gamma$, $X$, and $M$ high-symmetry points for the square lattice. Labels in (a) denote the ``Mott features''.
}
\end{center}
\end{figure}
\subsection{Overview}
Figure~\ref{fig:1} shows the calculated spectral function \Akw\ at two extreme dopings: undoped (half-filling) and heavily doped.  At half-filling, a large Mott gap ($\sim 4t$) separates the lower Hubbard band (LHB) and the upper Hubbard bands (UHB). Within each band, there are two main spectral features [see the markers in Fig.~\ref{fig:1}(a)]: one at low binding energies, describable within a spin-polaron framework\,\cite{Martinez:1991SpinPolaron, Efstratios:2007SCBA}; and a second at higher energies, which results from an effective intra-sublattice hopping\,\cite{wang2015origin,wang2018erratum}. A ``waterfall''-like step connects the two features, constituting one of the critical spectral signatures of correlation effects.\,\cite{ronning2005anomalous, macridin2007high, moritz2009effect, wang2015origin} Throughout this Letter, we refer to these single-particle these features that exist already at the half-filled Hubbard model as the \emph{Mott features}. In the other extreme doping limit [see Figs.~\ref{fig:1}(b) and (c)], the spectrum resembles that of non-interacting electrons, although the Hubbard $U$ {remains unchanged}.  A quasiparticle dispersion across $E_{\rm F}$ dominates the spectral function, with indiscernible residual spectral weight on the other side of the Mott gap. This feature, to which we refer as the \emph{quasiparticle}, follows the tight-binding functional form of the bare band structure, but is subject to a doping-dependent bandwidth renormalization\,\cite{Kohno:2012PRL, PhysRevB.90.035111, wang2018influence}.

\subsection{Density of States}
Between these two limits, we first investigate the density of states (DOS) evolution with doping [see Figs.~\ref{fig:2}(a-d)]. We see that a remnant Mott gap exists at all doping levels and well separates the UHB and LHB. Though it is somewhat expected in a single-band Hubbard model, we want to emphasize that the robustness of the charge-transfer gap in doped cuprates has bee recently confirmed through STM experiments \cite{zhong2020direct}. Moreover, infinitesimal doping leads to the development of spectral weight at $E_{\rm F}$ for both electron- and hole-doping. With increasing doping, the spectral weight transfer gradually depletes the upper (lower) Hubbard band, and the chemical potential smoothly evolves away from half-filling (with a finite linewidth, one can still define a $E_{\rm F}$ at $n\!=\!1$). In this process, the transferred spectral weight becomes energetically mixed with the lower (upper) Hubbard band upon doping, rather than forming an entirely separate in-gap state\,\cite{imada1998metal,shen2004missing}. 

To quantify such spectral weight evolution, we integrate the three regions: residual LHB and UHB (red and blue) and the doping-induced states (green), respectively. Unlike the separations at $E_{\rm F}$ in Figs.~\ref{fig:2}(a,d), we further assume that the residual LHB (UHB) in a hole-doped (electron-doped) system carries the same spectral weight as the UHB (LHB) and, therefore, assign a small portion of spectral weight below (above) $E_{\rm F}$ to the doping-induced states [see Fig.~\ref{fig:2}(c)]. We believe this assignment gives a better characterization compared to previous studies using small-cluster Hubbard or charge-transfer models\cite{eskes1991anomalous, dagotto1991density, chen1991electronic,  imada1998metal, liebsch2010spectral, phillips2010colloquium, Kohno:2012PRL}, because we have considered that the doping-induced feature (i.e. quasiparticle) penetrates $E_{\rm F}$ and coexists with the Mott features. As such, with the increasing doping, the rapid growth of it spectral weight involves contributions from both the LHB and UHB via doping. Our analysis shows that initially the spectral weight changes as $\sim2x$ (where $x$ is the concentration of doped carriers), but is reduced to $0.84x$ starting from roughly optimal doping. This gradual change reflects the unraveling of electronic correlations upon doping.

\subsection{Spectral Function}
Taking a closer look in momentum space to identify where and how the quasiparticles emerge via \Akw, Fig.~\ref{fig:3} shows the doping dependent single-particle spectra along high-symmetry cuts. In contrast to the back-bending spin-polaron at half-filling, doping immediately leads to the appearance of spectral weight near $E_{\rm F}$ [see Fig.~\ref{fig:3}(a)], with a heavily renormalized quasiparticle dispersion consistent with ARPES experiments on the underdoped cuprates\,\cite{yoshida2006systematic,shen2005nodal, shen2004missing, shen2007angle}.  The spectral weight of this quasiparticle grows monotonically with doping: it gradually ``fills-in'' near the $M$-point for hole doping, and the $\Gamma$-point for electron doping.  A clear suppression of the antinodal spectral weight at low hole doping reflects the \textit{pseudogap}, while a similar suppression of the node at low electron doping indicates the \textit{hotspot}  [see Supplementary Note 1]. The appearance of these two distinct phenomena suggests that the major normal-state quasiparticle features at optimal doping may also be qualitatively captured by a single-band Hubbard model.\cite{Damascelli2003, armitage2010progress,senechal2004hot} Upon further doping, the renormalization gradually decreases, and the quasiparticle dispersion smoothly evolves into the free dispersion shown in Figs.~\ref{fig:1}(b) and (c).

\begin{figure}[t!]
\begin{center}
\includegraphics[width=\columnwidth]{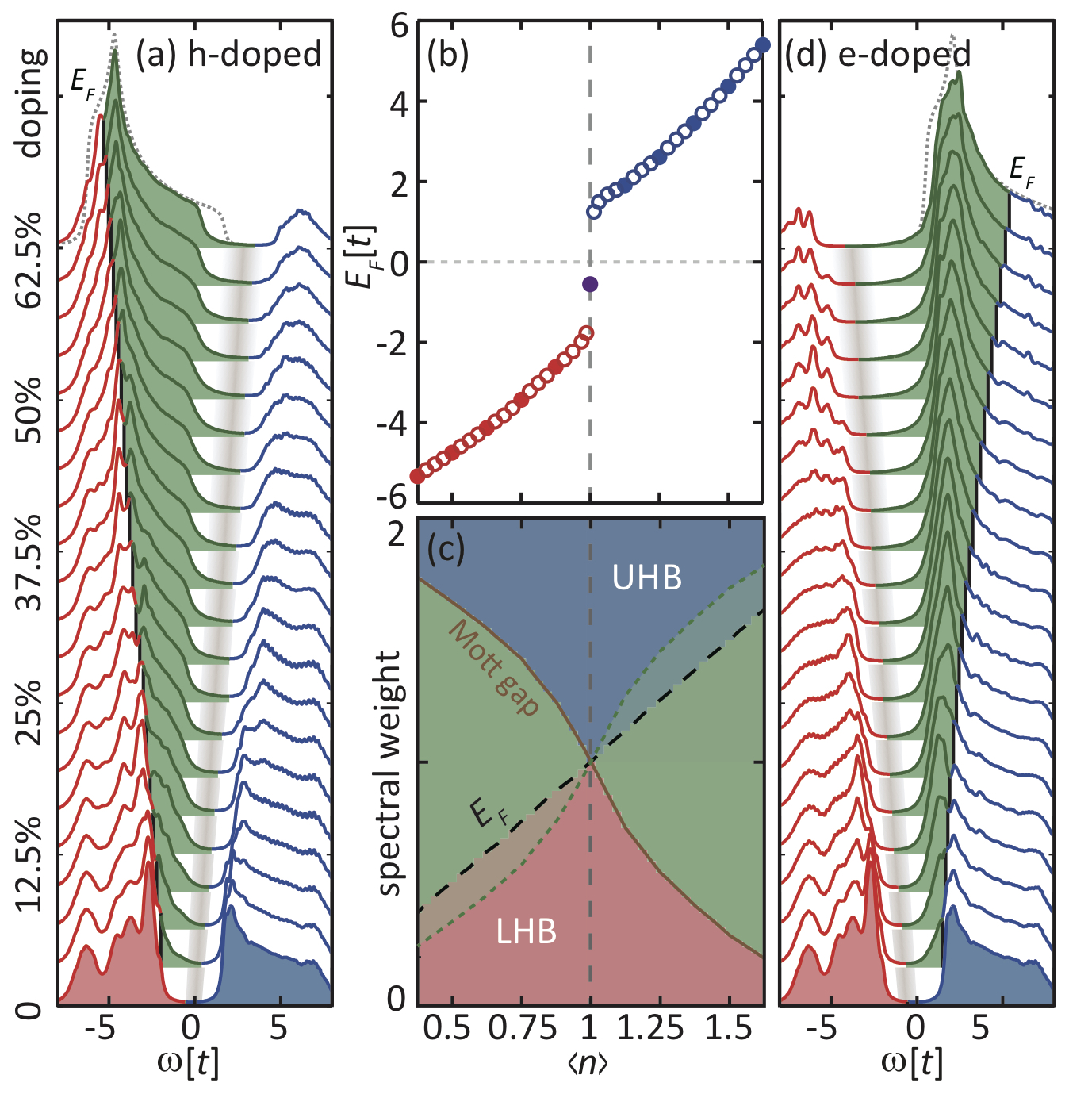}
\caption{\label{fig:2} {\bf Doping dependence of density of states. }
(a,d) Evolution of the calculated density of states (DOS) with hole and electron doping, respectively. The green regions highlight the doping-induced states above (hole-doping) and below (electron-doping) $E_{\rm F}$.  The grey dashed lines denote the DOS for the non-interacting model, and the brown shades indicate the Mott gap. (b) The dependence of $E_{\rm F}$ on the electron density $n$, reflecting the Mott plateau. The solid (open) circles show results from calculations without (with) a supercluster construction. (c) The integrated weight of panels (a) and (d).  The doping-induced and remnant spectral weight in the lower Hubbard band (LHB) and the upper Hubbard bands (UHB) are plotted in green, red and blue, respectively. The solid brown curve denotes their separation at the gap center, while the dashed green curve denotes the separation between doping-induced states and the LHB (UHB), assuming the latter has equal weight with the UHB (LHB).
}
\end{center}
\end{figure}

\begin{figure*}[!ht]
\begin{center}
\includegraphics[width=18cm]{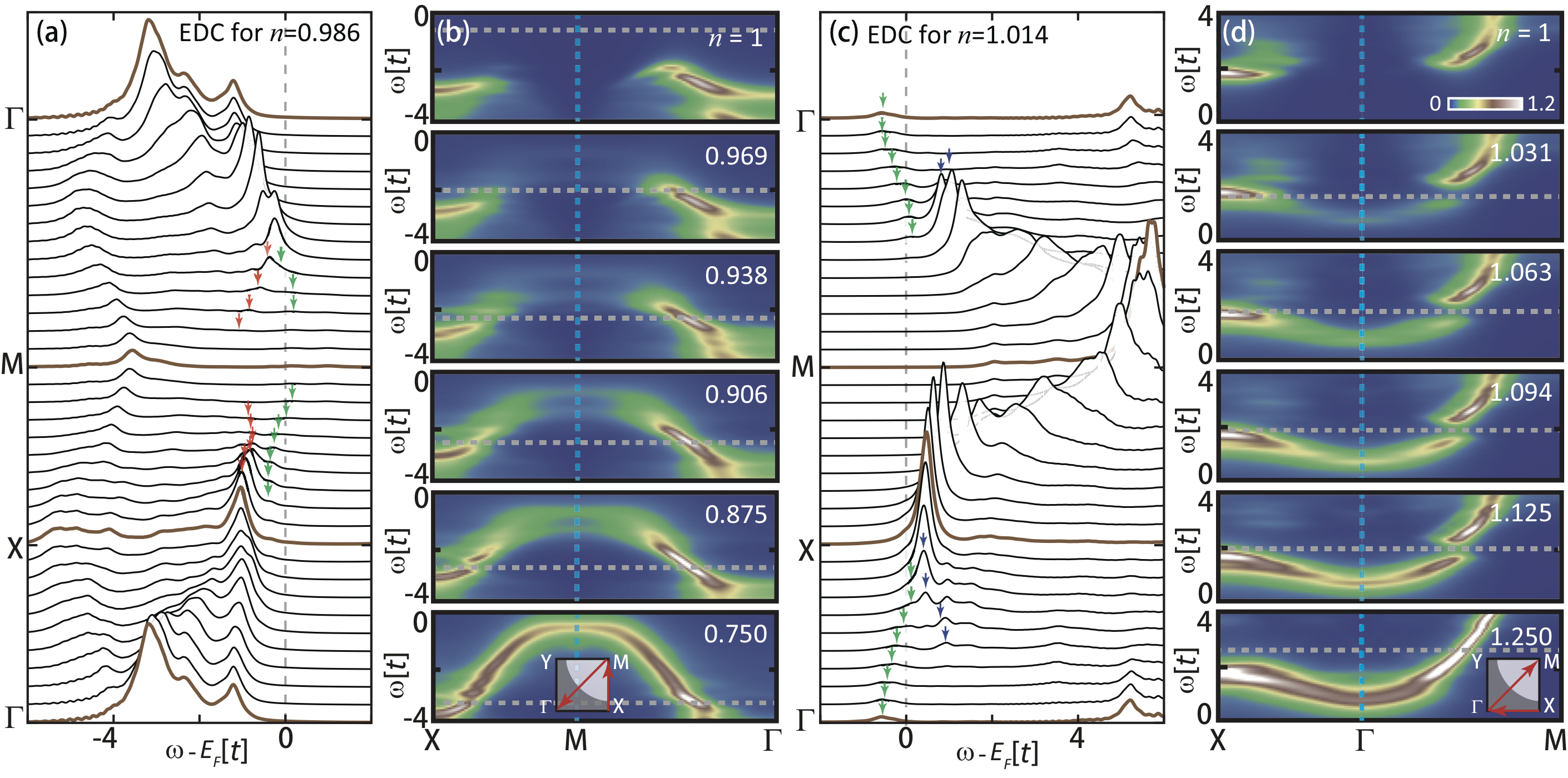}
\caption{\label{fig:3} {\bf Doping dependence of spectral functions. }
(a) Energy distribution curves (EDCs) for calculated spectral functions with $1.4\%$ hole doping. (b) False-color plot for the spectral function \Akw\ near $E_{\rm F}$ along high-symmetry cuts, as indicated by the inset, for a fixed energy window and various electron concentrations with hole doping. (c,d) Same as (a,b) but for electron doping instead. In (a) and (c), the green arrows denote the quasiparticle features, while the red and blue arrows denote the remnant spin-polaron features at the lower- and upper-Hubbard band, respectively. In all panels, the gray dashed line denotes Fermi level.  
}
\end{center}
\end{figure*}

The difference between the Hubbard model calculation and ARPES spectrum mainly lies in the lineshape near $k_F$. In contrast to a broad peak describable by the strong polaronic dressing at half-filling and light doping\,\cite{shen2004missing}, the renormalized quasiparticle in Fig.~\ref{fig:3} is always sharp near $(\pi/2,\pi/2)$ for hole doping [also see the Supplementary Note 2]. That means the linewidth change in the underdoped regime cannot be attributed simply to the quasiparticle dressed by spin excitations in the doped Hubbard model. Correlation-enhanced phonon polaronic dressing may be required to reproduce the experimental lineshape\,\cite{mishchenko2004electron}. The omission of the phonon dressing also may relate to the shape of the chemical potential jump $\sim t\sim$300meV in Fig.~\ref{fig:2}(b) for a small doping out of half-filling, which is observed smoother in cuprate experiments\,\cite{yagi2006chemical}.

\subsection{Doping Dependence of Spectral Weights}
\begin{figure}[t!]
\begin{center}
\includegraphics[width=\columnwidth]{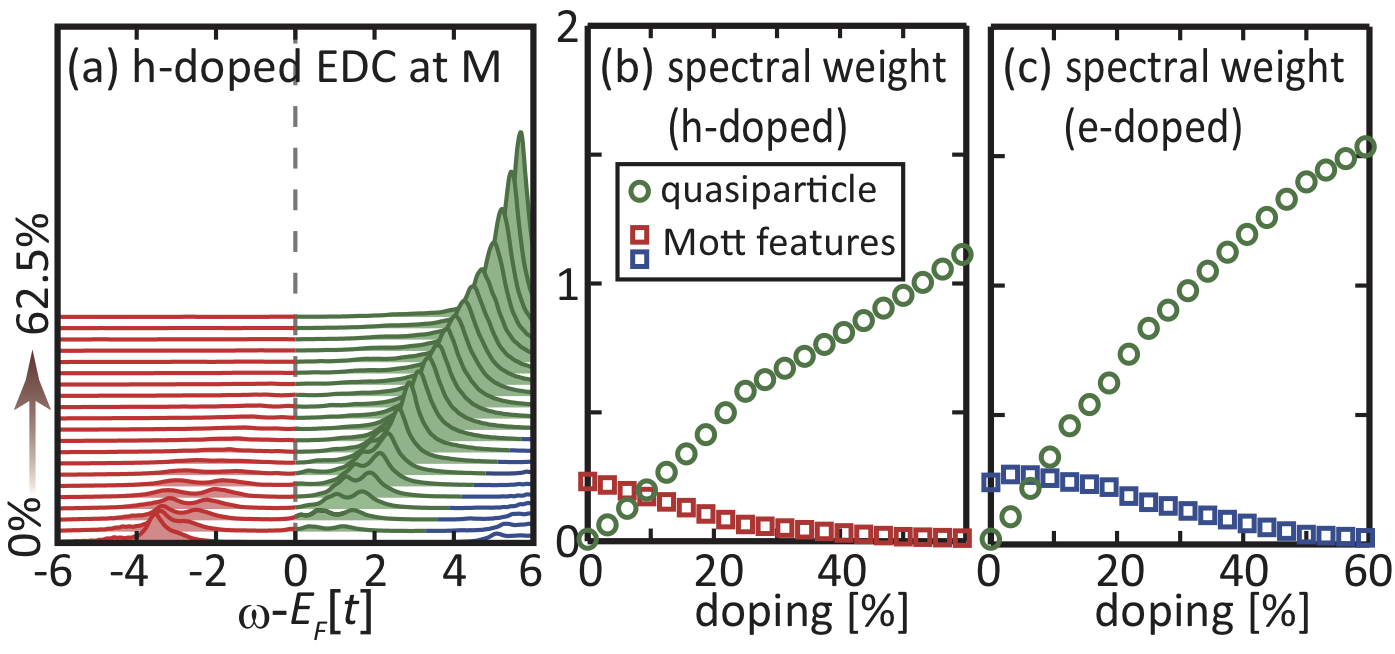}
\caption{\label{fig:4} {\bf Spectral weight evolution of well-separated features. }
(a) Energy distribution curves at the $M$-point for different hole doping levels. The green shaded region highlights the growth of the quasiparticle, while red marks residual Mott features in the lower Hubbard band. 
(b) Integrated spectral weight of features in (a). (c) A similar analysis performed in electron-doped systems at the $\Gamma$-point instead, where blue marks the corresponding integrated weight in the upper Hubbard bands.
}
\end{center}
\end{figure}

Besides the qualitative spectral shape, we perform quantitative analysis of the spectral weight doping evolution at two distinct momenta and within reprsentative energy ranges. We first focus on the $M$-point, where the quasiparticle is most separated from the higher-energy Mott features [see Fig.~\ref{fig:1} and Fig.~\ref{fig:3}(b)]. In Fig.~\ref{fig:4}(a), we observe rapid growth in the spectral weight associated with the quasiparticle (green), overwhelming the remnants of the lower Hubbard band (red) at a moderate hole concentration. Residual spectral weight of the latter features gradually decreases with doping until their disappearance at $\sim\! 20\%$ doping [see Fig.~\ref{fig:4}(b)]. The visibility of these Mott features in doped systems has two interesting implications. On one hand, the coupling between carriers and spin fluctuations is present even in a regime without long-range magnetic order, consistent with several recent experimental observations\,\cite{le2011intense, dean2013persistence, dean2013high, ishii2014high}. On the other hand, the eventual disappearance of these Mott features may account for the transition to a more metallic phase at $\sim 20$\% doping in Bi$_2$Sr$_2$CaCu$_2$O$_{8+x}$\,\cite{he2018rapid,chen2019strange} and YBa$_2$Cu$_3$O$_{6+x}$\,\cite{fournier2010loss}.

A similar analysis for electron doping, now at the $\Gamma$-point, indicates substantial particle-hole asymmetry with respect to doping [see Fig.~\ref{fig:4}(c)]. The Mott features persist until even higher doping on the electron-doped side. Although the finite cluster size of CPT calculations precludes definitive assessment of long-range order, the correlation effects at heavy doping suggest a substantial impact of spin correlations. This agrees with the recent discovery of Fermi surface reconstruction outside the AFM phase in Nd$_{2-x}$Ce$_x$CuO$_4$\,\cite{he2019fermi}. Here, the visibility of Mott features at higher doping (up to $\sim\!40\%$) than the hole-doped side reflects more robust magnetic correlations in the electron-doped Hubbard model\,\cite{tohyama2004asymmetry, moritz2011investigation, wang2014real}.

\begin{figure}[t!]
\begin{center}
\includegraphics[width=\columnwidth]{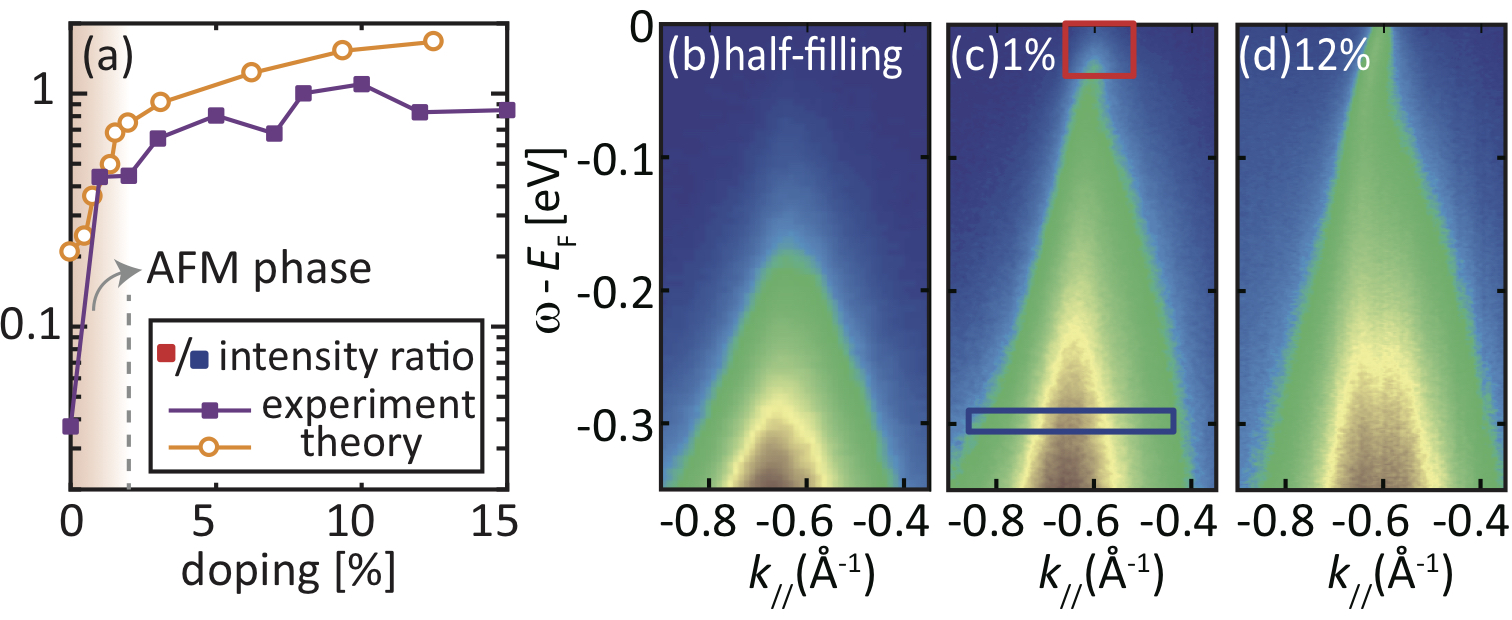}
\caption{\label{fig:5} {\bf Spectral weight evolution near $k_{\rm F}$. }
(a) The spectral weight ratio between the lower and higher energy windows obtained from calculations (open circles) and experiments (solid squares). The shaded region denotes the antiferromagnetic phase in (La,Sr)$_{2}$CuO$_{4}$. (b-d) Angle-resolved photoemission spectroscopy experiments on underdoped (La,Sr)$_{2}$CuO$_{4}$ for half-filling, 1\% and 12\% doping near $k_F$. The red and blue boxes in (c) denote the lower and higher energy windows near Fermi momentum $k_F$, corresponding to the quasiparticle and polaronic feature, respectively. Their ratios correspond to the dots in panel (a).}
\end{center}
\end{figure}

To verify the above dichotomy within occupied states measurable by ARPES experiments, we focus our attention on the spectrum near $k_{\rm F}$ along the nodal direction, where the dispersion crosses $E_{\rm F}$. Comparing the spectral weight in two momentum-energy windows -- one close to $E_{\rm F}$ and a second at higher binding energy -- we can quantitatively distinguish the evolution of the quasiparticle spectral weight from the Mott features (specifically spin-polaron in the calculation) at low doping.  As shown in Fig.~\ref{fig:5}(a), following a rapid exchange of spectral weight below $2\%$ hole doping, both features begin to saturate and coexist (with comparable spectral intensities).  We also observe consistent behavior in experimental spectra for the extremely underdoped regime of La$_{2-x}$Sr$_x$CuO$_4$ [see Figs.~\ref{fig:5}(b-d)]. After immediate development of the quasiparticle feature at 1\% doping, the spectral weight ratio between the quasiparticle and polaronic features saturates in the subsequent large range of doping. Such a saturation has been also observed in optimally doped HgBa$_2$CuO$_{4+\delta}$\cite{vishik2014angle} and Nd$_{2-x}$Ce$_x$CuO$_4$ \cite{he2019fermi}. This agreement confirms our theoretical prediction that strong correlations continue to play an essential role up to relatively high doping levels. [Note that the two integration windows have different areas: the upper one (red) covers $(k_F-0.05\AA, k_F+0.05\AA)\times(-0.05\textrm{eV},0)$, while the lower one (blue) is $(k_F-0.23\AA, k_F+0.07\AA)\times(-0.31\textrm{eV},-0.29\textrm{eV})$. Therefore, the ratio in Fig.~5(d) gives the trend for doping dependence instead of the absolute value of the quasiparticle weight.]

In contrast to the agreement of the spectral weight's doping dependence, the numerically calculated (nodal) spectral features of the Hubbard model are sharper than those found in the ARPES experiment of cuprates [see the comparison between Fig.~\ref{fig:3} and \ref{fig:5}]. While one culprit is the lack of the electron-phonon coupling in the calculations, we cannot completely rule out the possibility that the quasiparticle may become less coherent already within the Hubbard model---but only when explicitly calculated using far larger clusters (which is, as of now, unfeasible due to the exponential growth of the Hilbert space with the cluster size) [see the Methods section].

\section{Conclusion}
We have presented a comprehensive benchmark study of the single-particle spectral function of cuprates upon hole and electron doping, from the perspective of the single-band Hubbard model. We have analyzed and dissected the various features of the single-particle spectra, showing their distinct origins by carefully examining their doping dependences. Many of the conclusions drawn from the Hubbard model show good agreement with existing and here reported ARPES experiments. Starting from an extremely underdoped regime, doped carriers induce quasiparticle-like states near the Fermi level. Both the momentum-resolved calculations and ARPES experiments reveal that these itinerant states have no analog in the Mott insulating state at half-filling (i.e.~the Mott gap, spin-polaron, and high-energy intra-sublattice features). Instead, an important finding of this work is that these emergent features coexist with the original Mott features in a wide range of doping. Though heavily renormalized with small spectral weight at low doping, this quasiparticle feature gradual emerges from the Mott feature at a rate roughly twice that of the doping. With heavy doping ($\sim\!20\%$ for hole-doping), the continued presence and electron-hole asymmetry of correlations are consistent with the observations in recent ARPES and RIXS experiments\,\cite{chen2019strange, he2019fermi, he2018rapid, graf2007universal, le2011intense, dean2013persistence, dean2013high}.

From these aspects, the single-band Hubbard model seems to effectively capture the essence of the emergence of low energy quasiparticles, from the extremely underdoped regime to the overdoped regime. However, in contrast to these qualitative consistencies in spectral weight and dispersion, the almost doping-independent lineshape calculated from the Hubbard model cannot address the observations in parent compounds or lightly doped cuprates; the purely electronic Hubbard model also fails to reproduce the widely observed low-energy kinks in the quasiparticle bands\,\cite{bogdanov2000evidence, lanzara2001evidence, zhou2003high}. Towards a more comprehensive picture, including lattice polaronic coupling provides another channel to destroy the quasiparticles' coherence\cite{shen2004missing}, contributing to material-specific dependence in various cuprate families as well as other transition metal oxides.

\section{Methods}
\subsection{Hubbard Model}
The Hamiltonian of the single-band Hubbard model is given by\,\cite{Zhang:1988jf, Eskes:1988ef}
\begin{equation} \label{eq:hubbard}
\mathcal{H}=-\! \! \sum_{{\bf i},{\bf j},\sigma} \!\left(\!t_{\bf ij} c^\dagger_{{\bf j} \sigma} c_{{\bf i} \sigma} \!+\!h.c.\! \right) \!+ U\sum_{{\bf i}}\left(\!n_{{\bf i}\uparrow}\!-\!\frac12\!\right)\left(\!n_{{\bf i}\downarrow}\!-\!\frac12\!\right).
\end{equation}
Here, $c^\dagger_{{\bf i}\sigma}$ ($c_{{\bf i}\sigma}$) and $n_{{\bf i}\sigma}$ denote the creation (annihilation) and density operators at site $\textbf{i}$ of spin $\sigma$, respectively; $U$ denotes the on-site Coulomb interaction; and $t_{\bf ij}$ encodes the electron hopping, restricted here to nearest-neighbors $t_{\bf \langle ij\rangle}\!=\!t$ and next-nearest-neighbors $t_{\bf \langle\!\langle ij\rangle\!\rangle}\!=\!t^{\prime}$.  We chose parameters $U=8t$ and $t^\prime=-0.3t$, common for simulations of cuprates. 

\subsection{Single-Particle Spectral Function and Cluster Perturbation Theory}
The single particle spectral function, which corresponds to the ARPES measurement, is defined as:
\begin{equation}\label{eq:akw}
A({\bf k},\omega)=-\frac1{\pi}\textrm{Im} \sum_\sigma\langle G|c^\dagger_{{\bf k}\sigma}\frac1{\omega+\mathcal{H}-E_G+i\delta}c_{{\bf k} \sigma}|G\rangle,
\end{equation}
where $c_{{\bf k} \sigma} = \sum_{\bf i}c_{{\bf i} \sigma} e^{-i\textbf{k}\cdot\textbf{r}_\textbf{i}} /\sqrt{N}$ denotes the electron annihilation operator at momentum space, $|G\rangle$ is the ground state with energy $E_G$. We choose a Lorentzian energy broadening of $\delta=0.15t$ throughout this paper.

The idea of cluster perturbation theory (CPT) was invented by S{\'e}n{\'e}chal \emph{et al}.~\cite{Senechal:2000fg, Senechal:2002fr} and was demonstrated to be an efficient approach to estimate the \Akw\ of strongly correlated systems at zero temperature. For completeness, we briefly reiterate the CPT method here, while the readers should refer to Ref.~\onlinecite{Senechal:2002fr} for details. Assuming one divides the infinite plane into clusters, the Hamiltonian can be split into $\mathcal{H}=\mathcal{H}_c +\mathcal{H}_{\rm int}$. Here $\mathcal{H}_c$ contains the (open-boundary) intra-cluster operators, while $\mathcal{H}_{\rm int}$ contains the operators with inter-cluster indices (hopping terms for the Hubbard model). Then one can use exact diagonalization to precisely solve the Green's function $G_c(\omega)$ associated with the intra-cluster Hamiltonian $\mathcal{H}_{c}$. Here, we use the parallel Arnoldi method and the Paradeisos algorithm to determine the equilibrium ground-state wavefunction.\,\cite{lehoucq1998arpack,  jia2017paradeisos}

The CPT method then estimate the Green's function of the infinite plane by treating $\mathcal{H}_{\rm int}$ perturbatively, giving
\begin{equation}\label{eq:CPT1}
\mathcal{G}(\textbf{k},\omega)=\frac{G_c(\omega)}{1-{V}(\textbf{k})G_c(\omega)}\,.
\end{equation}
where ${V}(\textbf{k}) = \sum_\mathbf{R}\mathcal{H}_{\rm int} e^{i\textbf{k}\cdot\textbf{R}}$ is the inter-cluster interactions projected to the intra-cluster coordinates. Then taking the long-wavelength limit, we obtain
\begin{equation}
A(\textbf{k},\omega)_{\rm CPT}=-\frac1{\pi N}\textrm{Im}\sum_{a,b}\mathcal{G}_{a,b}(\textbf{k},\omega)e^{i\textbf{k}\cdot(\textbf{r}_a-\textbf{r}_b)},
\end{equation}
where $a,b$ are intra-cluster site indices. Although CPT gives corrections to the small-cluster ED calculations and reduces the finite-size effect, the finite-size effect does not entirely disappear in the CPT calculations\cite{Senechal:2000fg,Senechal:2002fr}. With an increase of the system size, the number of poles of the $G_c(\omega)$ grows [see Fig.~2 in Ref~\onlinecite{Senechal:2000fg}].

The CPT method can be further generalized into superclusters, giving access to any rational doping values. Denoting the Green's function of a (disconnected) supercluster consisting of $M$ clusters as $G_0=\oplus_m^M G_c^{(m)}$, where $G_c^{(m)}$ is the intra-cluster Green's function of the $m$-th cluster. Note that these $M$ clusters can be filled by different integer numbers, leading to an averaged rational filling in the supercluster. Then following the CPT philosophy, one can first include the inter-cluster perturbation inside the supercluster, giving
\begin{equation}
G_{sc} = G_0 (1-\mathcal{H}_{\rm int}^{(ic)} G_0)^{-1}
\end{equation}
where $\mathcal{H}_{\rm int}^{(ic)}$ is the inter-cluster Hamiltonian terms within the supercluster. The CPT correlation of the supercluster Green's function then gives
\begin{eqnarray}
&&A(\textbf{k},\omega)_{\rm SC-CPT}\nonumber\\
&=&-\frac1{\pi N}\textrm{Im}\sum_{a,b}\left[\frac{G_0(\omega)}{1-\left(\mathcal{H}_{\rm int}^{(ic)} + V_{sc}(\mathbf{k})\right)G_0(\omega)}\right]_{ab}\!e^{i\textbf{k}\cdot(\textbf{r}_a-\textbf{r}_b)}\,.
\end{eqnarray}
Similar to Eq.~\eqref{eq:CPT1}, the $V_{sc}(\mathbf{k})$ is the inter-supercluster hopping. In the calculations of the main text, we evaluate the $4\times4$ cluster spectral function by an exact solver and $8\times8$ or $12\times12$ by a supercluster solver.\cite{Senechal:2002fr} 

The density of states (DOS) is calculated through the integration of \Akw\ over the first Brillium zone:
\begin{eqnarray}
\mathrm{DOS}(\omega) = \int_{\rm BZ} \frac{d\textbf{k}^2}{4\pi^2} A(\textbf{k},\omega)_{\rm SC-CPT}\,.
\end{eqnarray}
The Fermi level is then determined by the equality $\mathrm{DOS}(E_{\rm F})=n$.

\subsection{Experimental details}
High-quality single crystals of La$_{2-x}$Sr$_x$CuO$_4$ (LSCO) are synthesized with traveling solvent floating zone method, and subsequently cut into $\sim$1$\times$1$\times$0.5 mm$^3$ pieces along the crystal axes. They are then glued to oxygen-free copper sample holders with alternating layers of conductive silver epoxy and Torr seal resin. Ceramic top posts are then glued to the surface of the single crystals, only to be cleaved under ultrahigh vacuum to expose fresh surfaces for photoemission measurements.

Angle-resolved photoemission spectroscopy on ten different dopings of LSCO is performed at beamline 10.0.1 in the Advanced Light Source at Lawrence Berkeley National Laboratory and beamline 5-4 at Stanford Synchrotron Radiation Lightsource at SLAC National Laboratory. The vacuum is kept at better than 5$\times$10$^{-11}$ Torr throughout the experiments. 55~eV and 20~eV photons are used with the linear polarization parallel to the diagonal Cu-Cu (nodal) direction. All data shown here are collected at 10-15~K except for $x$ = 0 sample, which is measured at 50~K to mitigate static charging effect.

\section*{Acknowledgements}
This work was supported by the U.S. Department of Energy, Office of Basic Energy Sciences, Division of Materials Sciences and Engineering, under Contract No.~DE-AC02-76SF00515. Y.W. acknowledges the National Science Foundation (NSF) through grant No.DMR-2038011. Y.H. acknowledges support from the Miller Institute for Basic Research in Science. K.W. acknowledges support from the Polish National Science Center (NCN) under Project No.~2016/22/E/ST3/00560 and 2016/23/B/ST3/00839. E.W.H. was supported by the Gordon and Betty Moore Foundation EPiQS Initiative through the grant GBMF 4305 and GBMF 8691. This research used resources of the National Energy Research Scientific Computing Center (NERSC), a U.S. Department of Energy Office of Science User Facility operated under Contract No. DE-AC02-05CH11231. The ARPES experiments were performed at the Advanced Light Source (ALS), which is a DOE Office of Science User Facility under Contract No. DE-AC02-05CH11231, and at Stanford Synchrotron Radiation Lightsource (SSRL) which is supported by the DOE, Office of Science, Basic Energy Sciences, Division of Materials Science and Engineering, under Contract No. DE-AC02-76SF00515.

\section*{Author contributions} 
Y.W. and K.W. performed the calculations. Y.H. carried out the ARPES measurements with support from M.H., D.L. and S.M. S.K. grew the single crystals used in the experiment. Y.W. and Y.H. performed the data analysis. B.M. Z.-X.S. and T.P.D. provided guidance to the designed research. Y.W., Y.H. and K.W. wrote the paper with the help from E.W.H., B.M., C.J. and T.P.D..

\bibliography{paper}

\begin{thebibliography}{91}%
\makeatletter
\providecommand \@ifxundefined [1]{%
 \@ifx{#1\undefined}
}%
\providecommand \@ifnum [1]{%
 \ifnum #1\expandafter \@firstoftwo
 \else \expandafter \@secondoftwo
 \fi
}%
\providecommand \@ifx [1]{%
 \ifx #1\expandafter \@firstoftwo
 \else \expandafter \@secondoftwo
 \fi
}%
\providecommand \natexlab [1]{#1}%
\providecommand \enquote  [1]{``#1''}%
\providecommand \bibnamefont  [1]{#1}%
\providecommand \bibfnamefont [1]{#1}%
\providecommand \citenamefont [1]{#1}%
\providecommand \href@noop [0]{\@secondoftwo}%
\providecommand \href [0]{\begingroup \@sanitize@url \@href}%
\providecommand \@href[1]{\@@startlink{#1}\@@href}%
\providecommand \@@href[1]{\endgroup#1\@@endlink}%
\providecommand \@sanitize@url [0]{\catcode `\\12\catcode `\$12\catcode
  `\&12\catcode `\#12\catcode `\^12\catcode `\_12\catcode `\%12\relax}%
\providecommand \@@startlink[1]{}%
\providecommand \@@endlink[0]{}%
\providecommand \url  [0]{\begingroup\@sanitize@url \@url }%
\providecommand \@url [1]{\endgroup\@href {#1}{\urlprefix }}%
\providecommand \urlprefix  [0]{URL }%
\providecommand \Eprint [0]{\href }%
\providecommand \doibase [0]{http://dx.doi.org/}%
\providecommand \selectlanguage [0]{\@gobble}%
\providecommand \bibinfo  [0]{\@secondoftwo}%
\providecommand \bibfield  [0]{\@secondoftwo}%
\providecommand \translation [1]{[#1]}%
\providecommand \BibitemOpen [0]{}%
\providecommand \bibitemStop [0]{}%
\providecommand \bibitemNoStop [0]{.\EOS\space}%
\providecommand \EOS [0]{\spacefactor3000\relax}%
\providecommand \BibitemShut  [1]{\csname bibitem#1\endcsname}%
\let\auto@bib@innerbib\@empty
\bibitem [{\citenamefont {Wen}\ and\ \citenamefont {Lee}(1996)}]{Wen1996}%
  \BibitemOpen
  \bibfield  {author} {\bibinfo {author} {\bibfnamefont {X.-G.}\ \bibnamefont
  {Wen}}\ and\ \bibinfo {author} {\bibfnamefont {P.~A.}\ \bibnamefont {Lee}},\
  }\href@noop {} {\bibfield  {journal} {\bibinfo  {journal} {Phys. Rev. Lett.}\
  }\textbf {\bibinfo {volume} {76}},\ \bibinfo {pages} {503} (\bibinfo {year}
  {1996})}\BibitemShut {NoStop}%
\bibitem [{\citenamefont {Lee}\ \emph {et~al.}(2006)\citenamefont {Lee},
  \citenamefont {Nagaosa},\ and\ \citenamefont {Wen}}]{lee2006doping}%
  \BibitemOpen
  \bibfield  {author} {\bibinfo {author} {\bibfnamefont {P.~A.}\ \bibnamefont
  {Lee}}, \bibinfo {author} {\bibfnamefont {N.}~\bibnamefont {Nagaosa}}, \ and\
  \bibinfo {author} {\bibfnamefont {X.-G.}\ \bibnamefont {Wen}},\ }\href@noop
  {} {\bibfield  {journal} {\bibinfo  {journal} {Rev. Mod. Phys.}\ }\textbf
  {\bibinfo {volume} {78}},\ \bibinfo {pages} {17} (\bibinfo {year}
  {2006})}\BibitemShut {NoStop}%
\bibitem [{\citenamefont {Stephan}\ and\ \citenamefont
  {Horsch}(1991)}]{Stephan1991}%
  \BibitemOpen
  \bibfield  {author} {\bibinfo {author} {\bibfnamefont {W.}~\bibnamefont
  {Stephan}}\ and\ \bibinfo {author} {\bibfnamefont {P.}~\bibnamefont
  {Horsch}},\ }\href@noop {} {\bibfield  {journal} {\bibinfo  {journal} {Phys.
  Rev. Lett.}\ }\textbf {\bibinfo {volume} {66}},\ \bibinfo {pages} {2258}
  (\bibinfo {year} {1991})}\BibitemShut {NoStop}%
\bibitem [{\citenamefont {Dagotto}(1994)}]{Dagotto:1994cz}%
  \BibitemOpen
  \bibfield  {author} {\bibinfo {author} {\bibfnamefont {E.}~\bibnamefont
  {Dagotto}},\ }\href@noop {} {\bibfield  {journal} {\bibinfo  {journal} {Rev.
  Mod. Phys.}\ }\textbf {\bibinfo {volume} {66}},\ \bibinfo {pages} {763}
  (\bibinfo {year} {1994})}\BibitemShut {NoStop}%
\bibitem [{\citenamefont {Damascelli}\ \emph {et~al.}(2003)\citenamefont
  {Damascelli}, \citenamefont {Hussain},\ and\ \citenamefont
  {Shen}}]{Damascelli2003}%
  \BibitemOpen
  \bibfield  {author} {\bibinfo {author} {\bibfnamefont {A.}~\bibnamefont
  {Damascelli}}, \bibinfo {author} {\bibfnamefont {Z.}~\bibnamefont {Hussain}},
  \ and\ \bibinfo {author} {\bibfnamefont {Z.-X.}\ \bibnamefont {Shen}},\
  }\href@noop {} {\bibfield  {journal} {\bibinfo  {journal} {Rev. Mod. Phys.}\
  }\textbf {\bibinfo {volume} {75}},\ \bibinfo {pages} {473} (\bibinfo {year}
  {2003})}\BibitemShut {NoStop}%
\bibitem [{\citenamefont {Zhang}\ and\ \citenamefont
  {Rice}(1988{\natexlab{a}})}]{zhang1988effective}%
  \BibitemOpen
  \bibfield  {author} {\bibinfo {author} {\bibfnamefont {F.}~\bibnamefont
  {Zhang}}\ and\ \bibinfo {author} {\bibfnamefont {T.}~\bibnamefont {Rice}},\
  }\href@noop {} {\bibfield  {journal} {\bibinfo  {journal} {Phys. Rev. B}\
  }\textbf {\bibinfo {volume} {37}},\ \bibinfo {pages} {3759} (\bibinfo {year}
  {1988}{\natexlab{a}})}\BibitemShut {NoStop}%
\bibitem [{\citenamefont {Ronning}\ \emph {et~al.}(2005)\citenamefont
  {Ronning}, \citenamefont {Shen}, \citenamefont {Armitage}, \citenamefont
  {Damascelli}, \citenamefont {Lu}, \citenamefont {Shen}, \citenamefont
  {Miller},\ and\ \citenamefont {Kim}}]{ronning2005anomalous}%
  \BibitemOpen
  \bibfield  {author} {\bibinfo {author} {\bibfnamefont {F.}~\bibnamefont
  {Ronning}}, \bibinfo {author} {\bibfnamefont {K.~M.}\ \bibnamefont {Shen}},
  \bibinfo {author} {\bibfnamefont {N.~P.}\ \bibnamefont {Armitage}}, \bibinfo
  {author} {\bibfnamefont {A.}~\bibnamefont {Damascelli}}, \bibinfo {author}
  {\bibfnamefont {D.~H.}\ \bibnamefont {Lu}}, \bibinfo {author} {\bibfnamefont
  {Z.-X.}\ \bibnamefont {Shen}}, \bibinfo {author} {\bibfnamefont {L.~L.}\
  \bibnamefont {Miller}}, \ and\ \bibinfo {author} {\bibfnamefont
  {C.}~\bibnamefont {Kim}},\ }\href@noop {} {\bibfield  {journal} {\bibinfo
  {journal} {Phys. Rev. B}\ }\textbf {\bibinfo {volume} {71}},\ \bibinfo
  {pages} {094518} (\bibinfo {year} {2005})}\BibitemShut {NoStop}%
\bibitem [{\citenamefont {Meevasana}\ \emph {et~al.}(2007)\citenamefont
  {Meevasana}, \citenamefont {Zhou}, \citenamefont {Sahrakorpi}, \citenamefont
  {Lee}, \citenamefont {Yang}, \citenamefont {Tanaka}, \citenamefont
  {Mannella}, \citenamefont {Yoshida}, \citenamefont {Lu}, \citenamefont {Chen}
  \emph {et~al.}}]{meevasana2007hierarchy}%
  \BibitemOpen
  \bibfield  {author} {\bibinfo {author} {\bibfnamefont {W.}~\bibnamefont
  {Meevasana}}, \bibinfo {author} {\bibfnamefont {X.}~\bibnamefont {Zhou}},
  \bibinfo {author} {\bibfnamefont {S.}~\bibnamefont {Sahrakorpi}}, \bibinfo
  {author} {\bibfnamefont {W.}~\bibnamefont {Lee}}, \bibinfo {author}
  {\bibfnamefont {W.}~\bibnamefont {Yang}}, \bibinfo {author} {\bibfnamefont
  {K.}~\bibnamefont {Tanaka}}, \bibinfo {author} {\bibfnamefont
  {N.}~\bibnamefont {Mannella}}, \bibinfo {author} {\bibfnamefont
  {T.}~\bibnamefont {Yoshida}}, \bibinfo {author} {\bibfnamefont
  {D.}~\bibnamefont {Lu}}, \bibinfo {author} {\bibfnamefont {Y.}~\bibnamefont
  {Chen}},  \emph {et~al.},\ }\href@noop {} {\bibfield  {journal} {\bibinfo
  {journal} {Phys. Rev. B}\ }\textbf {\bibinfo {volume} {75}},\ \bibinfo
  {pages} {174506} (\bibinfo {year} {2007})}\BibitemShut {NoStop}%
\bibitem [{\citenamefont {Valla}\ \emph {et~al.}(2007)\citenamefont {Valla},
  \citenamefont {Kidd}, \citenamefont {Yin}, \citenamefont {Gu}, \citenamefont
  {Johnson}, \citenamefont {Pan},\ and\ \citenamefont
  {Fedorov}}]{valla2007high}%
  \BibitemOpen
  \bibfield  {author} {\bibinfo {author} {\bibfnamefont {T.}~\bibnamefont
  {Valla}}, \bibinfo {author} {\bibfnamefont {T.}~\bibnamefont {Kidd}},
  \bibinfo {author} {\bibfnamefont {W.-G.}\ \bibnamefont {Yin}}, \bibinfo
  {author} {\bibfnamefont {G.}~\bibnamefont {Gu}}, \bibinfo {author}
  {\bibfnamefont {P.}~\bibnamefont {Johnson}}, \bibinfo {author} {\bibfnamefont
  {Z.-H.}\ \bibnamefont {Pan}}, \ and\ \bibinfo {author} {\bibfnamefont
  {A.}~\bibnamefont {Fedorov}},\ }\href@noop {} {\bibfield  {journal} {\bibinfo
   {journal} {Phys. Rev. Lett.}\ }\textbf {\bibinfo {volume} {98}},\ \bibinfo
  {pages} {167003} (\bibinfo {year} {2007})}\BibitemShut {NoStop}%
\bibitem [{\citenamefont {Xie}\ \emph {et~al.}(2007)\citenamefont {Xie},
  \citenamefont {Yang}, \citenamefont {Shen}, \citenamefont {Zhao},
  \citenamefont {Ou}, \citenamefont {Wei}, \citenamefont {Gu}, \citenamefont
  {Arita}, \citenamefont {Qiao}, \citenamefont {Namatame} \emph
  {et~al.}}]{xie2007high}%
  \BibitemOpen
  \bibfield  {author} {\bibinfo {author} {\bibfnamefont {B.}~\bibnamefont
  {Xie}}, \bibinfo {author} {\bibfnamefont {K.}~\bibnamefont {Yang}}, \bibinfo
  {author} {\bibfnamefont {D.}~\bibnamefont {Shen}}, \bibinfo {author}
  {\bibfnamefont {J.}~\bibnamefont {Zhao}}, \bibinfo {author} {\bibfnamefont
  {H.}~\bibnamefont {Ou}}, \bibinfo {author} {\bibfnamefont {J.}~\bibnamefont
  {Wei}}, \bibinfo {author} {\bibfnamefont {S.}~\bibnamefont {Gu}}, \bibinfo
  {author} {\bibfnamefont {M.}~\bibnamefont {Arita}}, \bibinfo {author}
  {\bibfnamefont {S.}~\bibnamefont {Qiao}}, \bibinfo {author} {\bibfnamefont
  {H.}~\bibnamefont {Namatame}},  \emph {et~al.},\ }\href@noop {} {\bibfield
  {journal} {\bibinfo  {journal} {Phys. Rev. Lett.}\ }\textbf {\bibinfo
  {volume} {98}},\ \bibinfo {pages} {147001} (\bibinfo {year}
  {2007})}\BibitemShut {NoStop}%
\bibitem [{\citenamefont {Martinez}\ and\ \citenamefont
  {Horsch}(1991)}]{Martinez:1991SpinPolaron}%
  \BibitemOpen
  \bibfield  {author} {\bibinfo {author} {\bibfnamefont {G.}~\bibnamefont
  {Martinez}}\ and\ \bibinfo {author} {\bibfnamefont {P.}~\bibnamefont
  {Horsch}},\ }\href@noop {} {\bibfield  {journal} {\bibinfo  {journal} {Phys.
  Rev. B}\ }\textbf {\bibinfo {volume} {44}},\ \bibinfo {pages} {317} (\bibinfo
  {year} {1991})}\BibitemShut {NoStop}%
\bibitem [{\citenamefont {Ba{\l}a}\ \emph {et~al.}(1995)\citenamefont
  {Ba{\l}a}, \citenamefont {Ole{\'s}},\ and\ \citenamefont
  {Zaanen}}]{Bala:1995ttprJSCBA}%
  \BibitemOpen
  \bibfield  {author} {\bibinfo {author} {\bibfnamefont {J.}~\bibnamefont
  {Ba{\l}a}}, \bibinfo {author} {\bibfnamefont {A.}~\bibnamefont {Ole{\'s}}}, \
  and\ \bibinfo {author} {\bibfnamefont {J.}~\bibnamefont {Zaanen}},\
  }\href@noop {} {\bibfield  {journal} {\bibinfo  {journal} {Phys. Rev. B}\
  }\textbf {\bibinfo {volume} {52}},\ \bibinfo {pages} {4597} (\bibinfo {year}
  {1995})}\BibitemShut {NoStop}%
\bibitem [{\citenamefont {Macridin}\ \emph {et~al.}(2007)\citenamefont
  {Macridin}, \citenamefont {Jarrell}, \citenamefont {Maier},\ and\
  \citenamefont {Scalapino}}]{macridin2007high}%
  \BibitemOpen
  \bibfield  {author} {\bibinfo {author} {\bibfnamefont {A.}~\bibnamefont
  {Macridin}}, \bibinfo {author} {\bibfnamefont {M.}~\bibnamefont {Jarrell}},
  \bibinfo {author} {\bibfnamefont {T.}~\bibnamefont {Maier}}, \ and\ \bibinfo
  {author} {\bibfnamefont {D.}~\bibnamefont {Scalapino}},\ }\href@noop {}
  {\bibfield  {journal} {\bibinfo  {journal} {Phys. Rev. Lett.}\ }\textbf
  {\bibinfo {volume} {99}},\ \bibinfo {pages} {237001} (\bibinfo {year}
  {2007})}\BibitemShut {NoStop}%
\bibitem [{\citenamefont {Manousakis}(2007)}]{Efstratios:2007SCBA}%
  \BibitemOpen
  \bibfield  {author} {\bibinfo {author} {\bibfnamefont {E.}~\bibnamefont
  {Manousakis}},\ }\href@noop {} {\bibfield  {journal} {\bibinfo  {journal}
  {Phys. Rev. B}\ }\textbf {\bibinfo {volume} {75}},\ \bibinfo {pages} {035106}
  (\bibinfo {year} {2007})}\BibitemShut {NoStop}%
\bibitem [{\citenamefont {Graf}\ \emph {et~al.}(2007)\citenamefont {Graf},
  \citenamefont {Gweon}, \citenamefont {McElroy}, \citenamefont {Zhou},
  \citenamefont {Jozwiak}, \citenamefont {Rotenberg}, \citenamefont {Bill},
  \citenamefont {Sasagawa}, \citenamefont {Eisaki}, \citenamefont {Uchida}
  \emph {et~al.}}]{graf2007universal}%
  \BibitemOpen
  \bibfield  {author} {\bibinfo {author} {\bibfnamefont {J.}~\bibnamefont
  {Graf}}, \bibinfo {author} {\bibfnamefont {G.-H.}\ \bibnamefont {Gweon}},
  \bibinfo {author} {\bibfnamefont {K.}~\bibnamefont {McElroy}}, \bibinfo
  {author} {\bibfnamefont {S.}~\bibnamefont {Zhou}}, \bibinfo {author}
  {\bibfnamefont {C.}~\bibnamefont {Jozwiak}}, \bibinfo {author} {\bibfnamefont
  {E.}~\bibnamefont {Rotenberg}}, \bibinfo {author} {\bibfnamefont
  {A.}~\bibnamefont {Bill}}, \bibinfo {author} {\bibfnamefont {T.}~\bibnamefont
  {Sasagawa}}, \bibinfo {author} {\bibfnamefont {H.}~\bibnamefont {Eisaki}},
  \bibinfo {author} {\bibfnamefont {S.}~\bibnamefont {Uchida}},  \emph
  {et~al.},\ }\href@noop {} {\bibfield  {journal} {\bibinfo  {journal} {Phys.
  Rev. Lett.}\ }\textbf {\bibinfo {volume} {98}},\ \bibinfo {pages} {067004}
  (\bibinfo {year} {2007})}\BibitemShut {NoStop}%
\bibitem [{\citenamefont {He}\ \emph {et~al.}(2019)\citenamefont {He},
  \citenamefont {Rotundu}, \citenamefont {Scheurer}, \citenamefont {He},
  \citenamefont {Hashimoto}, \citenamefont {Xu}, \citenamefont {Wang},
  \citenamefont {Huang}, \citenamefont {Jia}, \citenamefont {Chen} \emph
  {et~al.}}]{he2019fermi}%
  \BibitemOpen
  \bibfield  {author} {\bibinfo {author} {\bibfnamefont {J.}~\bibnamefont
  {He}}, \bibinfo {author} {\bibfnamefont {C.~R.}\ \bibnamefont {Rotundu}},
  \bibinfo {author} {\bibfnamefont {M.~S.}\ \bibnamefont {Scheurer}}, \bibinfo
  {author} {\bibfnamefont {Y.}~\bibnamefont {He}}, \bibinfo {author}
  {\bibfnamefont {M.}~\bibnamefont {Hashimoto}}, \bibinfo {author}
  {\bibfnamefont {K.-J.}\ \bibnamefont {Xu}}, \bibinfo {author} {\bibfnamefont
  {Y.}~\bibnamefont {Wang}}, \bibinfo {author} {\bibfnamefont {E.~W.}\
  \bibnamefont {Huang}}, \bibinfo {author} {\bibfnamefont {T.}~\bibnamefont
  {Jia}}, \bibinfo {author} {\bibfnamefont {S.}~\bibnamefont {Chen}},  \emph
  {et~al.},\ }\href@noop {} {\bibfield  {journal} {\bibinfo  {journal} {Proc.
  Natl. Acad. Sci. U. S. A.}\ }\textbf {\bibinfo {volume} {116}},\ \bibinfo
  {pages} {3449} (\bibinfo {year} {2019})}\BibitemShut {NoStop}%
\bibitem [{\citenamefont {He}\ \emph {et~al.}(2018)\citenamefont {He},
  \citenamefont {Hashimoto}, \citenamefont {Song}, \citenamefont {Chen},
  \citenamefont {He}, \citenamefont {Vishik}, \citenamefont {Moritz},
  \citenamefont {Lee}, \citenamefont {Nagaosa}, \citenamefont {Zaanen} \emph
  {et~al.}}]{he2018rapid}%
  \BibitemOpen
  \bibfield  {author} {\bibinfo {author} {\bibfnamefont {Y.}~\bibnamefont
  {He}}, \bibinfo {author} {\bibfnamefont {M.}~\bibnamefont {Hashimoto}},
  \bibinfo {author} {\bibfnamefont {D.}~\bibnamefont {Song}}, \bibinfo {author}
  {\bibfnamefont {S.-D.}\ \bibnamefont {Chen}}, \bibinfo {author}
  {\bibfnamefont {J.}~\bibnamefont {He}}, \bibinfo {author} {\bibfnamefont
  {I.}~\bibnamefont {Vishik}}, \bibinfo {author} {\bibfnamefont
  {B.}~\bibnamefont {Moritz}}, \bibinfo {author} {\bibfnamefont {D.-H.}\
  \bibnamefont {Lee}}, \bibinfo {author} {\bibfnamefont {N.}~\bibnamefont
  {Nagaosa}}, \bibinfo {author} {\bibfnamefont {J.}~\bibnamefont {Zaanen}},
  \emph {et~al.},\ }\href@noop {} {\bibfield  {journal} {\bibinfo  {journal}
  {Science}\ }\textbf {\bibinfo {volume} {362}},\ \bibinfo {pages} {62}
  (\bibinfo {year} {2018})}\BibitemShut {NoStop}%
\bibitem [{\citenamefont {Chen}\ \emph {et~al.}(2019)\citenamefont {Chen},
  \citenamefont {Hashimoto}, \citenamefont {He}, \citenamefont {Song},
  \citenamefont {Xu}, \citenamefont {He}, \citenamefont {Devereaux},
  \citenamefont {Eisaki}, \citenamefont {Lu}, \citenamefont {Zaanen} \emph
  {et~al.}}]{chen2019strange}%
  \BibitemOpen
  \bibfield  {author} {\bibinfo {author} {\bibfnamefont {S.-D.}\ \bibnamefont
  {Chen}}, \bibinfo {author} {\bibfnamefont {M.}~\bibnamefont {Hashimoto}},
  \bibinfo {author} {\bibfnamefont {Y.}~\bibnamefont {He}}, \bibinfo {author}
  {\bibfnamefont {D.}~\bibnamefont {Song}}, \bibinfo {author} {\bibfnamefont
  {K.-J.}\ \bibnamefont {Xu}}, \bibinfo {author} {\bibfnamefont {J.-F.}\
  \bibnamefont {He}}, \bibinfo {author} {\bibfnamefont {T.~P.}\ \bibnamefont
  {Devereaux}}, \bibinfo {author} {\bibfnamefont {H.}~\bibnamefont {Eisaki}},
  \bibinfo {author} {\bibfnamefont {D.-H.}\ \bibnamefont {Lu}}, \bibinfo
  {author} {\bibfnamefont {J.}~\bibnamefont {Zaanen}},  \emph {et~al.},\
  }\href@noop {} {\bibfield  {journal} {\bibinfo  {journal} {Science}\ }\textbf
  {\bibinfo {volume} {366}},\ \bibinfo {pages} {1099} (\bibinfo {year}
  {2019})}\BibitemShut {NoStop}%
\bibitem [{\citenamefont {Le~Tacon}\ \emph {et~al.}(2011)\citenamefont
  {Le~Tacon}, \citenamefont {Ghiringhelli}, \citenamefont {Chaloupka},
  \citenamefont {Sala}, \citenamefont {Hinkov}, \citenamefont {Haverkort},
  \citenamefont {Minola}, \citenamefont {Bakr}, \citenamefont {Zhou},
  \citenamefont {Blanco-Canosa} \emph {et~al.}}]{le2011intense}%
  \BibitemOpen
  \bibfield  {author} {\bibinfo {author} {\bibfnamefont {M.}~\bibnamefont
  {Le~Tacon}}, \bibinfo {author} {\bibfnamefont {G.}~\bibnamefont
  {Ghiringhelli}}, \bibinfo {author} {\bibfnamefont {J.}~\bibnamefont
  {Chaloupka}}, \bibinfo {author} {\bibfnamefont {M.~M.}\ \bibnamefont {Sala}},
  \bibinfo {author} {\bibfnamefont {V.}~\bibnamefont {Hinkov}}, \bibinfo
  {author} {\bibfnamefont {M.}~\bibnamefont {Haverkort}}, \bibinfo {author}
  {\bibfnamefont {M.}~\bibnamefont {Minola}}, \bibinfo {author} {\bibfnamefont
  {M.}~\bibnamefont {Bakr}}, \bibinfo {author} {\bibfnamefont {K.}~\bibnamefont
  {Zhou}}, \bibinfo {author} {\bibfnamefont {S.}~\bibnamefont {Blanco-Canosa}},
   \emph {et~al.},\ }\href@noop {} {\bibfield  {journal} {\bibinfo  {journal}
  {Nat. Phys.}\ }\textbf {\bibinfo {volume} {7}},\ \bibinfo {pages} {725}
  (\bibinfo {year} {2011})}\BibitemShut {NoStop}%
\bibitem [{\citenamefont {Dean}\ \emph
  {et~al.}(2013{\natexlab{a}})\citenamefont {Dean}, \citenamefont {Dellea},
  \citenamefont {Springell}, \citenamefont {Yakhou-Harris}, \citenamefont
  {Kummer}, \citenamefont {Brookes}, \citenamefont {Liu}, \citenamefont {Sun},
  \citenamefont {Strle}, \citenamefont {Schmitt} \emph
  {et~al.}}]{dean2013persistence}%
  \BibitemOpen
  \bibfield  {author} {\bibinfo {author} {\bibfnamefont {M.}~\bibnamefont
  {Dean}}, \bibinfo {author} {\bibfnamefont {G.}~\bibnamefont {Dellea}},
  \bibinfo {author} {\bibfnamefont {R.}~\bibnamefont {Springell}}, \bibinfo
  {author} {\bibfnamefont {F.}~\bibnamefont {Yakhou-Harris}}, \bibinfo {author}
  {\bibfnamefont {K.}~\bibnamefont {Kummer}}, \bibinfo {author} {\bibfnamefont
  {N.}~\bibnamefont {Brookes}}, \bibinfo {author} {\bibfnamefont
  {X.}~\bibnamefont {Liu}}, \bibinfo {author} {\bibfnamefont {Y.}~\bibnamefont
  {Sun}}, \bibinfo {author} {\bibfnamefont {J.}~\bibnamefont {Strle}}, \bibinfo
  {author} {\bibfnamefont {T.}~\bibnamefont {Schmitt}},  \emph {et~al.},\
  }\href@noop {} {\bibfield  {journal} {\bibinfo  {journal} {Nat. Mater.}\
  }\textbf {\bibinfo {volume} {12}},\ \bibinfo {pages} {1019} (\bibinfo {year}
  {2013}{\natexlab{a}})}\BibitemShut {NoStop}%
\bibitem [{\citenamefont {Dean}\ \emph
  {et~al.}(2013{\natexlab{b}})\citenamefont {Dean}, \citenamefont {James},
  \citenamefont {Springell}, \citenamefont {Liu}, \citenamefont {Monney},
  \citenamefont {Zhou}, \citenamefont {Konik}, \citenamefont {Wen},
  \citenamefont {Xu}, \citenamefont {Gu} \emph {et~al.}}]{dean2013high}%
  \BibitemOpen
  \bibfield  {author} {\bibinfo {author} {\bibfnamefont {M.}~\bibnamefont
  {Dean}}, \bibinfo {author} {\bibfnamefont {A.}~\bibnamefont {James}},
  \bibinfo {author} {\bibfnamefont {R.}~\bibnamefont {Springell}}, \bibinfo
  {author} {\bibfnamefont {X.}~\bibnamefont {Liu}}, \bibinfo {author}
  {\bibfnamefont {C.}~\bibnamefont {Monney}}, \bibinfo {author} {\bibfnamefont
  {K.}~\bibnamefont {Zhou}}, \bibinfo {author} {\bibfnamefont {R.}~\bibnamefont
  {Konik}}, \bibinfo {author} {\bibfnamefont {J.}~\bibnamefont {Wen}}, \bibinfo
  {author} {\bibfnamefont {Z.}~\bibnamefont {Xu}}, \bibinfo {author}
  {\bibfnamefont {G.}~\bibnamefont {Gu}},  \emph {et~al.},\ }\href@noop {}
  {\bibfield  {journal} {\bibinfo  {journal} {Phys. Rev. Lett.}\ }\textbf
  {\bibinfo {volume} {110}},\ \bibinfo {pages} {147001} (\bibinfo {year}
  {2013}{\natexlab{b}})}\BibitemShut {NoStop}%
\bibitem [{\citenamefont {Lee}\ \emph {et~al.}(2014)\citenamefont {Lee},
  \citenamefont {Lee}, \citenamefont {Nowadnick}, \citenamefont {Gerber},
  \citenamefont {Tabis}, \citenamefont {Huang}, \citenamefont {Strocov},
  \citenamefont {Motoyama}, \citenamefont {Yu}, \citenamefont {Moritz} \emph
  {et~al.}}]{lee2014asymmetry}%
  \BibitemOpen
  \bibfield  {author} {\bibinfo {author} {\bibfnamefont {W.}~\bibnamefont
  {Lee}}, \bibinfo {author} {\bibfnamefont {J.}~\bibnamefont {Lee}}, \bibinfo
  {author} {\bibfnamefont {E.}~\bibnamefont {Nowadnick}}, \bibinfo {author}
  {\bibfnamefont {S.}~\bibnamefont {Gerber}}, \bibinfo {author} {\bibfnamefont
  {W.}~\bibnamefont {Tabis}}, \bibinfo {author} {\bibfnamefont
  {S.}~\bibnamefont {Huang}}, \bibinfo {author} {\bibfnamefont
  {V.}~\bibnamefont {Strocov}}, \bibinfo {author} {\bibfnamefont
  {E.}~\bibnamefont {Motoyama}}, \bibinfo {author} {\bibfnamefont
  {G.}~\bibnamefont {Yu}}, \bibinfo {author} {\bibfnamefont {B.}~\bibnamefont
  {Moritz}},  \emph {et~al.},\ }\href@noop {} {\bibfield  {journal} {\bibinfo
  {journal} {Nat. Phys.}\ } (\bibinfo {year} {2014})}\BibitemShut {NoStop}%
\bibitem [{\citenamefont {Ishii}\ \emph {et~al.}(2014)\citenamefont {Ishii},
  \citenamefont {Fujita}, \citenamefont {Sasaki}, \citenamefont {Minola},
  \citenamefont {Dellea}, \citenamefont {Mazzoli}, \citenamefont {Kummer},
  \citenamefont {Ghiringhelli}, \citenamefont {Braicovich}, \citenamefont
  {Tohyama} \emph {et~al.}}]{ishii2014high}%
  \BibitemOpen
  \bibfield  {author} {\bibinfo {author} {\bibfnamefont {K.}~\bibnamefont
  {Ishii}}, \bibinfo {author} {\bibfnamefont {M.}~\bibnamefont {Fujita}},
  \bibinfo {author} {\bibfnamefont {T.}~\bibnamefont {Sasaki}}, \bibinfo
  {author} {\bibfnamefont {M.}~\bibnamefont {Minola}}, \bibinfo {author}
  {\bibfnamefont {G.}~\bibnamefont {Dellea}}, \bibinfo {author} {\bibfnamefont
  {C.}~\bibnamefont {Mazzoli}}, \bibinfo {author} {\bibfnamefont
  {K.}~\bibnamefont {Kummer}}, \bibinfo {author} {\bibfnamefont
  {G.}~\bibnamefont {Ghiringhelli}}, \bibinfo {author} {\bibfnamefont
  {L.}~\bibnamefont {Braicovich}}, \bibinfo {author} {\bibfnamefont
  {T.}~\bibnamefont {Tohyama}},  \emph {et~al.},\ }\href@noop {} {\bibfield
  {journal} {\bibinfo  {journal} {Nat. Commun.}\ }\textbf {\bibinfo {volume}
  {5}} (\bibinfo {year} {2014})}\BibitemShut {NoStop}%
\bibitem [{\citenamefont {Shen}\ \emph {et~al.}(2005)\citenamefont {Shen},
  \citenamefont {Ronning}, \citenamefont {Lu}, \citenamefont {Baumberger},
  \citenamefont {Ingle}, \citenamefont {Lee}, \citenamefont {Meevasana},
  \citenamefont {Kohsaka}, \citenamefont {Azuma}, \citenamefont {Takano} \emph
  {et~al.}}]{shen2005nodal}%
  \BibitemOpen
  \bibfield  {author} {\bibinfo {author} {\bibfnamefont {K.~M.}\ \bibnamefont
  {Shen}}, \bibinfo {author} {\bibfnamefont {F.}~\bibnamefont {Ronning}},
  \bibinfo {author} {\bibfnamefont {D.}~\bibnamefont {Lu}}, \bibinfo {author}
  {\bibfnamefont {F.}~\bibnamefont {Baumberger}}, \bibinfo {author}
  {\bibfnamefont {N.}~\bibnamefont {Ingle}}, \bibinfo {author} {\bibfnamefont
  {W.}~\bibnamefont {Lee}}, \bibinfo {author} {\bibfnamefont {W.}~\bibnamefont
  {Meevasana}}, \bibinfo {author} {\bibfnamefont {Y.}~\bibnamefont {Kohsaka}},
  \bibinfo {author} {\bibfnamefont {M.}~\bibnamefont {Azuma}}, \bibinfo
  {author} {\bibfnamefont {M.}~\bibnamefont {Takano}},  \emph {et~al.},\
  }\href@noop {} {\bibfield  {journal} {\bibinfo  {journal} {Science}\ }\textbf
  {\bibinfo {volume} {307}},\ \bibinfo {pages} {901} (\bibinfo {year}
  {2005})}\BibitemShut {NoStop}%
\bibitem [{\citenamefont {Shen}\ \emph {et~al.}(2004)\citenamefont {Shen},
  \citenamefont {Ronning}, \citenamefont {Lu}, \citenamefont {Lee},
  \citenamefont {Ingle}, \citenamefont {Meevasana}, \citenamefont {Baumberger},
  \citenamefont {Damascelli}, \citenamefont {Armitage}, \citenamefont {Miller}
  \emph {et~al.}}]{shen2004missing}%
  \BibitemOpen
  \bibfield  {author} {\bibinfo {author} {\bibfnamefont {K.}~\bibnamefont
  {Shen}}, \bibinfo {author} {\bibfnamefont {F.}~\bibnamefont {Ronning}},
  \bibinfo {author} {\bibfnamefont {D.}~\bibnamefont {Lu}}, \bibinfo {author}
  {\bibfnamefont {W.}~\bibnamefont {Lee}}, \bibinfo {author} {\bibfnamefont
  {N.}~\bibnamefont {Ingle}}, \bibinfo {author} {\bibfnamefont
  {W.}~\bibnamefont {Meevasana}}, \bibinfo {author} {\bibfnamefont
  {F.}~\bibnamefont {Baumberger}}, \bibinfo {author} {\bibfnamefont
  {A.}~\bibnamefont {Damascelli}}, \bibinfo {author} {\bibfnamefont
  {N.}~\bibnamefont {Armitage}}, \bibinfo {author} {\bibfnamefont
  {L.}~\bibnamefont {Miller}},  \emph {et~al.},\ }\href@noop {} {\bibfield
  {journal} {\bibinfo  {journal} {Phys. Rev. Lett.}\ }\textbf {\bibinfo
  {volume} {93}},\ \bibinfo {pages} {267002} (\bibinfo {year}
  {2004})}\BibitemShut {NoStop}%
\bibitem [{\citenamefont {Mishchenko}\ and\ \citenamefont
  {Nagaosa}(2004)}]{mishchenko2004electron}%
  \BibitemOpen
  \bibfield  {author} {\bibinfo {author} {\bibfnamefont {A.}~\bibnamefont
  {Mishchenko}}\ and\ \bibinfo {author} {\bibfnamefont {N.}~\bibnamefont
  {Nagaosa}},\ }\href@noop {} {\bibfield  {journal} {\bibinfo  {journal} {Phys.
  Rev. Lett.}\ }\textbf {\bibinfo {volume} {93}},\ \bibinfo {pages} {036402}
  (\bibinfo {year} {2004})}\BibitemShut {NoStop}%
\bibitem [{\citenamefont {Mishchenko}\ and\ \citenamefont
  {Nagaosa}(2006)}]{mishchenko2006numerical}%
  \BibitemOpen
  \bibfield  {author} {\bibinfo {author} {\bibfnamefont {A.}~\bibnamefont
  {Mishchenko}}\ and\ \bibinfo {author} {\bibfnamefont {N.}~\bibnamefont
  {Nagaosa}},\ }\href@noop {} {\bibfield  {journal} {\bibinfo  {journal} {Phys.
  Rev. B}\ }\textbf {\bibinfo {volume} {73}},\ \bibinfo {pages} {092502}
  (\bibinfo {year} {2006})}\BibitemShut {NoStop}%
\bibitem [{\citenamefont {Shen}\ \emph {et~al.}(2007)\citenamefont {Shen},
  \citenamefont {Ronning}, \citenamefont {Meevasana}, \citenamefont {Lu},
  \citenamefont {Ingle}, \citenamefont {Baumberger}, \citenamefont {Lee},
  \citenamefont {Miller}, \citenamefont {Kohsaka}, \citenamefont {Azuma} \emph
  {et~al.}}]{shen2007angle}%
  \BibitemOpen
  \bibfield  {author} {\bibinfo {author} {\bibfnamefont {K.}~\bibnamefont
  {Shen}}, \bibinfo {author} {\bibfnamefont {F.}~\bibnamefont {Ronning}},
  \bibinfo {author} {\bibfnamefont {W.}~\bibnamefont {Meevasana}}, \bibinfo
  {author} {\bibfnamefont {D.}~\bibnamefont {Lu}}, \bibinfo {author}
  {\bibfnamefont {N.}~\bibnamefont {Ingle}}, \bibinfo {author} {\bibfnamefont
  {F.}~\bibnamefont {Baumberger}}, \bibinfo {author} {\bibfnamefont
  {W.}~\bibnamefont {Lee}}, \bibinfo {author} {\bibfnamefont {L.}~\bibnamefont
  {Miller}}, \bibinfo {author} {\bibfnamefont {Y.}~\bibnamefont {Kohsaka}},
  \bibinfo {author} {\bibfnamefont {M.}~\bibnamefont {Azuma}},  \emph
  {et~al.},\ }\href@noop {} {\bibfield  {journal} {\bibinfo  {journal} {Phys.
  Rev. B}\ }\textbf {\bibinfo {volume} {75}},\ \bibinfo {pages} {075115}
  (\bibinfo {year} {2007})}\BibitemShut {NoStop}%
\bibitem [{\citenamefont {R{\"o}sch}\ \emph {et~al.}(2005)\citenamefont
  {R{\"o}sch}, \citenamefont {Gunnarsson}, \citenamefont {Zhou}, \citenamefont
  {Yoshida}, \citenamefont {Sasagawa}, \citenamefont {Fujimori}, \citenamefont
  {Hussain}, \citenamefont {Shen},\ and\ \citenamefont
  {Uchida}}]{rosch2005polaronic}%
  \BibitemOpen
  \bibfield  {author} {\bibinfo {author} {\bibfnamefont {O.}~\bibnamefont
  {R{\"o}sch}}, \bibinfo {author} {\bibfnamefont {O.}~\bibnamefont
  {Gunnarsson}}, \bibinfo {author} {\bibfnamefont {X.}~\bibnamefont {Zhou}},
  \bibinfo {author} {\bibfnamefont {T.}~\bibnamefont {Yoshida}}, \bibinfo
  {author} {\bibfnamefont {T.}~\bibnamefont {Sasagawa}}, \bibinfo {author}
  {\bibfnamefont {A.}~\bibnamefont {Fujimori}}, \bibinfo {author}
  {\bibfnamefont {Z.}~\bibnamefont {Hussain}}, \bibinfo {author} {\bibfnamefont
  {Z.-X.}\ \bibnamefont {Shen}}, \ and\ \bibinfo {author} {\bibfnamefont
  {S.}~\bibnamefont {Uchida}},\ }\href@noop {} {\bibfield  {journal} {\bibinfo
  {journal} {Phys. Rev. Lett.}\ }\textbf {\bibinfo {volume} {95}},\ \bibinfo
  {pages} {227002} (\bibinfo {year} {2005})}\BibitemShut {NoStop}%
\bibitem [{\citenamefont {Slezak}\ \emph {et~al.}(2006)\citenamefont {Slezak},
  \citenamefont {Macridin}, \citenamefont {Sawatzky}, \citenamefont {Jarrell},\
  and\ \citenamefont {Maier}}]{slezak2006spectral}%
  \BibitemOpen
  \bibfield  {author} {\bibinfo {author} {\bibfnamefont {C.}~\bibnamefont
  {Slezak}}, \bibinfo {author} {\bibfnamefont {A.}~\bibnamefont {Macridin}},
  \bibinfo {author} {\bibfnamefont {G.}~\bibnamefont {Sawatzky}}, \bibinfo
  {author} {\bibfnamefont {M.}~\bibnamefont {Jarrell}}, \ and\ \bibinfo
  {author} {\bibfnamefont {T.~A.}\ \bibnamefont {Maier}},\ }\href@noop {}
  {\bibfield  {journal} {\bibinfo  {journal} {Phys. Rev. B}\ }\textbf {\bibinfo
  {volume} {73}},\ \bibinfo {pages} {205122} (\bibinfo {year}
  {2006})}\BibitemShut {NoStop}%
\bibitem [{\citenamefont {Gunnarsson}\ and\ \citenamefont
  {R{\"o}sch}(2006)}]{gunnarsson2006electron}%
  \BibitemOpen
  \bibfield  {author} {\bibinfo {author} {\bibfnamefont {O.}~\bibnamefont
  {Gunnarsson}}\ and\ \bibinfo {author} {\bibfnamefont {O.}~\bibnamefont
  {R{\"o}sch}},\ }\href@noop {} {\bibfield  {journal} {\bibinfo  {journal}
  {Phys. Rev. B}\ }\textbf {\bibinfo {volume} {73}},\ \bibinfo {pages} {174521}
  (\bibinfo {year} {2006})}\BibitemShut {NoStop}%
\bibitem [{\citenamefont {Tranquada}\ \emph {et~al.}(1994)\citenamefont
  {Tranquada}, \citenamefont {Buttrey}, \citenamefont {Sachan},\ and\
  \citenamefont {Lorenzo}}]{tranquada1994simultaneous}%
  \BibitemOpen
  \bibfield  {author} {\bibinfo {author} {\bibfnamefont {J.~M.}\ \bibnamefont
  {Tranquada}}, \bibinfo {author} {\bibfnamefont {D.}~\bibnamefont {Buttrey}},
  \bibinfo {author} {\bibfnamefont {V.}~\bibnamefont {Sachan}}, \ and\ \bibinfo
  {author} {\bibfnamefont {J.}~\bibnamefont {Lorenzo}},\ }\href@noop {}
  {\bibfield  {journal} {\bibinfo  {journal} {Phys. Rev. Lett.}\ }\textbf
  {\bibinfo {volume} {73}},\ \bibinfo {pages} {1003} (\bibinfo {year}
  {1994})}\BibitemShut {NoStop}%
\bibitem [{\citenamefont {Ramirez}\ \emph {et~al.}(1996)\citenamefont
  {Ramirez}, \citenamefont {Schiffer}, \citenamefont {Cheong}, \citenamefont
  {Chen}, \citenamefont {Bao}, \citenamefont {Palstra}, \citenamefont {Gammel},
  \citenamefont {Bishop},\ and\ \citenamefont
  {Zegarski}}]{ramirez1996thermodynamic}%
  \BibitemOpen
  \bibfield  {author} {\bibinfo {author} {\bibfnamefont {A.}~\bibnamefont
  {Ramirez}}, \bibinfo {author} {\bibfnamefont {P.}~\bibnamefont {Schiffer}},
  \bibinfo {author} {\bibfnamefont {S.-W.}\ \bibnamefont {Cheong}}, \bibinfo
  {author} {\bibfnamefont {C.}~\bibnamefont {Chen}}, \bibinfo {author}
  {\bibfnamefont {W.}~\bibnamefont {Bao}}, \bibinfo {author} {\bibfnamefont
  {T.}~\bibnamefont {Palstra}}, \bibinfo {author} {\bibfnamefont
  {P.}~\bibnamefont {Gammel}}, \bibinfo {author} {\bibfnamefont
  {D.}~\bibnamefont {Bishop}}, \ and\ \bibinfo {author} {\bibfnamefont
  {B.}~\bibnamefont {Zegarski}},\ }\href@noop {} {\bibfield  {journal}
  {\bibinfo  {journal} {Phys. Rev. Lett.}\ }\textbf {\bibinfo {volume} {76}},\
  \bibinfo {pages} {3188} (\bibinfo {year} {1996})}\BibitemShut {NoStop}%
\bibitem [{\citenamefont {Cwik}\ \emph {et~al.}(2009)\citenamefont {Cwik},
  \citenamefont {Benomar}, \citenamefont {Finger}, \citenamefont {Sidis},
  \citenamefont {Senff}, \citenamefont {Reuther}, \citenamefont {Lorenz},\ and\
  \citenamefont {Braden}}]{cwik2009magnetic}%
  \BibitemOpen
  \bibfield  {author} {\bibinfo {author} {\bibfnamefont {M.}~\bibnamefont
  {Cwik}}, \bibinfo {author} {\bibfnamefont {M.}~\bibnamefont {Benomar}},
  \bibinfo {author} {\bibfnamefont {T.}~\bibnamefont {Finger}}, \bibinfo
  {author} {\bibfnamefont {Y.}~\bibnamefont {Sidis}}, \bibinfo {author}
  {\bibfnamefont {D.}~\bibnamefont {Senff}}, \bibinfo {author} {\bibfnamefont
  {M.}~\bibnamefont {Reuther}}, \bibinfo {author} {\bibfnamefont
  {T.}~\bibnamefont {Lorenz}}, \ and\ \bibinfo {author} {\bibfnamefont
  {M.}~\bibnamefont {Braden}},\ }\href@noop {} {\bibfield  {journal} {\bibinfo
  {journal} {Phys. Rev. Lett.}\ }\textbf {\bibinfo {volume} {102}},\ \bibinfo
  {pages} {057201} (\bibinfo {year} {2009})}\BibitemShut {NoStop}%
\bibitem [{\citenamefont {Ulbrich}\ and\ \citenamefont
  {Braden}(2012)}]{ulbrich2012neutron}%
  \BibitemOpen
  \bibfield  {author} {\bibinfo {author} {\bibfnamefont {H.}~\bibnamefont
  {Ulbrich}}\ and\ \bibinfo {author} {\bibfnamefont {M.}~\bibnamefont
  {Braden}},\ }\href@noop {} {\bibfield  {journal} {\bibinfo  {journal}
  {Physica C}\ }\textbf {\bibinfo {volume} {481}},\ \bibinfo {pages} {31}
  (\bibinfo {year} {2012})}\BibitemShut {NoStop}%
\bibitem [{\citenamefont {Tranquada}(2013)}]{tranquada2013spins}%
  \BibitemOpen
  \bibfield  {author} {\bibinfo {author} {\bibfnamefont {J.~M.}\ \bibnamefont
  {Tranquada}},\ }in\ \href@noop {} {\emph {\bibinfo {booktitle} {AIP
  Conference Proceedings}}},\ Vol.\ \bibinfo {volume} {1550}\ (\bibinfo
  {organization} {AIP},\ \bibinfo {year} {2013})\ pp.\ \bibinfo {pages}
  {114--187}\BibitemShut {NoStop}%
\bibitem [{\citenamefont {Bogdanov}\ \emph {et~al.}(2000)\citenamefont
  {Bogdanov}, \citenamefont {Lanzara}, \citenamefont {Kellar}, \citenamefont
  {Zhou}, \citenamefont {Lu}, \citenamefont {Zheng}, \citenamefont {Gu},
  \citenamefont {Shimoyama}, \citenamefont {Kishio}, \citenamefont {Ikeda}
  \emph {et~al.}}]{bogdanov2000evidence}%
  \BibitemOpen
  \bibfield  {author} {\bibinfo {author} {\bibfnamefont {P.}~\bibnamefont
  {Bogdanov}}, \bibinfo {author} {\bibfnamefont {A.}~\bibnamefont {Lanzara}},
  \bibinfo {author} {\bibfnamefont {S.}~\bibnamefont {Kellar}}, \bibinfo
  {author} {\bibfnamefont {X.}~\bibnamefont {Zhou}}, \bibinfo {author}
  {\bibfnamefont {E.}~\bibnamefont {Lu}}, \bibinfo {author} {\bibfnamefont
  {W.}~\bibnamefont {Zheng}}, \bibinfo {author} {\bibfnamefont
  {G.}~\bibnamefont {Gu}}, \bibinfo {author} {\bibfnamefont {J.-I.}\
  \bibnamefont {Shimoyama}}, \bibinfo {author} {\bibfnamefont {K.}~\bibnamefont
  {Kishio}}, \bibinfo {author} {\bibfnamefont {H.}~\bibnamefont {Ikeda}},
  \emph {et~al.},\ }\href@noop {} {\bibfield  {journal} {\bibinfo  {journal}
  {Phys. Rev. Lett.}\ }\textbf {\bibinfo {volume} {85}},\ \bibinfo {pages}
  {2581} (\bibinfo {year} {2000})}\BibitemShut {NoStop}%
\bibitem [{\citenamefont {Lanzara}\ \emph {et~al.}(2001)\citenamefont
  {Lanzara}, \citenamefont {Bogdanov}, \citenamefont {Zhou}, \citenamefont
  {Kellar}, \citenamefont {Feng}, \citenamefont {Lu}, \citenamefont {Yoshida},
  \citenamefont {Eisaki}, \citenamefont {Fujimori}, \citenamefont {Kishio}
  \emph {et~al.}}]{lanzara2001evidence}%
  \BibitemOpen
  \bibfield  {author} {\bibinfo {author} {\bibfnamefont {A.}~\bibnamefont
  {Lanzara}}, \bibinfo {author} {\bibfnamefont {P.}~\bibnamefont {Bogdanov}},
  \bibinfo {author} {\bibfnamefont {X.}~\bibnamefont {Zhou}}, \bibinfo {author}
  {\bibfnamefont {S.}~\bibnamefont {Kellar}}, \bibinfo {author} {\bibfnamefont
  {D.}~\bibnamefont {Feng}}, \bibinfo {author} {\bibfnamefont {E.}~\bibnamefont
  {Lu}}, \bibinfo {author} {\bibfnamefont {T.}~\bibnamefont {Yoshida}},
  \bibinfo {author} {\bibfnamefont {H.}~\bibnamefont {Eisaki}}, \bibinfo
  {author} {\bibfnamefont {A.}~\bibnamefont {Fujimori}}, \bibinfo {author}
  {\bibfnamefont {K.}~\bibnamefont {Kishio}},  \emph {et~al.},\ }\href@noop {}
  {\bibfield  {journal} {\bibinfo  {journal} {Nature}\ }\textbf {\bibinfo
  {volume} {412}},\ \bibinfo {pages} {510} (\bibinfo {year}
  {2001})}\BibitemShut {NoStop}%
\bibitem [{\citenamefont {Zhou}\ \emph {et~al.}(2003)\citenamefont {Zhou},
  \citenamefont {Yoshida}, \citenamefont {Lanzara}, \citenamefont {Bogdanov},
  \citenamefont {Kellar}, \citenamefont {Shen}, \citenamefont {Yang},
  \citenamefont {Ronning}, \citenamefont {Sasagawa}, \citenamefont {Kakeshita}
  \emph {et~al.}}]{zhou2003high}%
  \BibitemOpen
  \bibfield  {author} {\bibinfo {author} {\bibfnamefont {X.~.~J.}\ \bibnamefont
  {Zhou}}, \bibinfo {author} {\bibfnamefont {T.}~\bibnamefont {Yoshida}},
  \bibinfo {author} {\bibfnamefont {A.}~\bibnamefont {Lanzara}}, \bibinfo
  {author} {\bibfnamefont {P.}~\bibnamefont {Bogdanov}}, \bibinfo {author}
  {\bibfnamefont {S.}~\bibnamefont {Kellar}}, \bibinfo {author} {\bibfnamefont
  {K.}~\bibnamefont {Shen}}, \bibinfo {author} {\bibfnamefont {W.}~\bibnamefont
  {Yang}}, \bibinfo {author} {\bibfnamefont {F.}~\bibnamefont {Ronning}},
  \bibinfo {author} {\bibfnamefont {T.}~\bibnamefont {Sasagawa}}, \bibinfo
  {author} {\bibfnamefont {T.}~\bibnamefont {Kakeshita}},  \emph {et~al.},\
  }\href@noop {} {\bibfield  {journal} {\bibinfo  {journal} {Nature}\ }\textbf
  {\bibinfo {volume} {423}},\ \bibinfo {pages} {398} (\bibinfo {year}
  {2003})}\BibitemShut {NoStop}%
\bibitem [{\citenamefont {Cuk}\ \emph {et~al.}(2004)\citenamefont {Cuk},
  \citenamefont {Baumberger}, \citenamefont {Lu}, \citenamefont {Ingle},
  \citenamefont {Zhou}, \citenamefont {Eisaki}, \citenamefont {Kaneko},
  \citenamefont {Hussain}, \citenamefont {Devereaux}, \citenamefont {Nagaosa}
  \emph {et~al.}}]{cuk2004coupling}%
  \BibitemOpen
  \bibfield  {author} {\bibinfo {author} {\bibfnamefont {T.}~\bibnamefont
  {Cuk}}, \bibinfo {author} {\bibfnamefont {F.}~\bibnamefont {Baumberger}},
  \bibinfo {author} {\bibfnamefont {D.}~\bibnamefont {Lu}}, \bibinfo {author}
  {\bibfnamefont {N.}~\bibnamefont {Ingle}}, \bibinfo {author} {\bibfnamefont
  {X.}~\bibnamefont {Zhou}}, \bibinfo {author} {\bibfnamefont {H.}~\bibnamefont
  {Eisaki}}, \bibinfo {author} {\bibfnamefont {N.}~\bibnamefont {Kaneko}},
  \bibinfo {author} {\bibfnamefont {Z.}~\bibnamefont {Hussain}}, \bibinfo
  {author} {\bibfnamefont {T.}~\bibnamefont {Devereaux}}, \bibinfo {author}
  {\bibfnamefont {N.}~\bibnamefont {Nagaosa}},  \emph {et~al.},\ }\href@noop {}
  {\bibfield  {journal} {\bibinfo  {journal} {Phys. Rev. Lett.}\ }\textbf
  {\bibinfo {volume} {93}},\ \bibinfo {pages} {117003} (\bibinfo {year}
  {2004})}\BibitemShut {NoStop}%
\bibitem [{\citenamefont {Johnston}\ \emph {et~al.}(2012)\citenamefont
  {Johnston}, \citenamefont {Vishik}, \citenamefont {Lee}, \citenamefont
  {Schmitt}, \citenamefont {Uchida}, \citenamefont {Fujita}, \citenamefont
  {Ishida}, \citenamefont {Nagaosa}, \citenamefont {Shen},\ and\ \citenamefont
  {Devereaux}}]{johnston2012evidence}%
  \BibitemOpen
  \bibfield  {author} {\bibinfo {author} {\bibfnamefont {S.}~\bibnamefont
  {Johnston}}, \bibinfo {author} {\bibfnamefont {I.}~\bibnamefont {Vishik}},
  \bibinfo {author} {\bibfnamefont {W.}~\bibnamefont {Lee}}, \bibinfo {author}
  {\bibfnamefont {F.}~\bibnamefont {Schmitt}}, \bibinfo {author} {\bibfnamefont
  {S.}~\bibnamefont {Uchida}}, \bibinfo {author} {\bibfnamefont
  {K.}~\bibnamefont {Fujita}}, \bibinfo {author} {\bibfnamefont
  {S.}~\bibnamefont {Ishida}}, \bibinfo {author} {\bibfnamefont
  {N.}~\bibnamefont {Nagaosa}}, \bibinfo {author} {\bibfnamefont
  {Z.}~\bibnamefont {Shen}}, \ and\ \bibinfo {author} {\bibfnamefont
  {T.}~\bibnamefont {Devereaux}},\ }\href@noop {} {\bibfield  {journal}
  {\bibinfo  {journal} {Phys. Rev. Lett.}\ }\textbf {\bibinfo {volume} {108}},\
  \bibinfo {pages} {166404} (\bibinfo {year} {2012})}\BibitemShut {NoStop}%
\bibitem [{\citenamefont {Ohta}\ \emph {et~al.}(1991)\citenamefont {Ohta},
  \citenamefont {Tohyama},\ and\ \citenamefont {Maekawa}}]{ohta1991apex}%
  \BibitemOpen
  \bibfield  {author} {\bibinfo {author} {\bibfnamefont {Y.}~\bibnamefont
  {Ohta}}, \bibinfo {author} {\bibfnamefont {T.}~\bibnamefont {Tohyama}}, \
  and\ \bibinfo {author} {\bibfnamefont {S.}~\bibnamefont {Maekawa}},\
  }\href@noop {} {\bibfield  {journal} {\bibinfo  {journal} {Phys. Rev. B}\
  }\textbf {\bibinfo {volume} {43}},\ \bibinfo {pages} {2968} (\bibinfo {year}
  {1991})}\BibitemShut {NoStop}%
\bibitem [{\citenamefont {Peng}\ \emph {et~al.}(2017)\citenamefont {Peng},
  \citenamefont {Dellea}, \citenamefont {Minola}, \citenamefont {Conni},
  \citenamefont {Amorese}, \citenamefont {Di~Castro}, \citenamefont {De~Luca},
  \citenamefont {Kummer}, \citenamefont {Salluzzo}, \citenamefont {Sun} \emph
  {et~al.}}]{peng2017influence}%
  \BibitemOpen
  \bibfield  {author} {\bibinfo {author} {\bibfnamefont {Y.}~\bibnamefont
  {Peng}}, \bibinfo {author} {\bibfnamefont {G.}~\bibnamefont {Dellea}},
  \bibinfo {author} {\bibfnamefont {M.}~\bibnamefont {Minola}}, \bibinfo
  {author} {\bibfnamefont {M.}~\bibnamefont {Conni}}, \bibinfo {author}
  {\bibfnamefont {A.}~\bibnamefont {Amorese}}, \bibinfo {author} {\bibfnamefont
  {D.}~\bibnamefont {Di~Castro}}, \bibinfo {author} {\bibfnamefont
  {G.}~\bibnamefont {De~Luca}}, \bibinfo {author} {\bibfnamefont
  {K.}~\bibnamefont {Kummer}}, \bibinfo {author} {\bibfnamefont
  {M.}~\bibnamefont {Salluzzo}}, \bibinfo {author} {\bibfnamefont
  {X.}~\bibnamefont {Sun}},  \emph {et~al.},\ }\href@noop {} {\bibfield
  {journal} {\bibinfo  {journal} {Nat. Phys.}\ }\textbf {\bibinfo {volume}
  {13}},\ \bibinfo {pages} {1201} (\bibinfo {year} {2017})}\BibitemShut
  {NoStop}%
\bibitem [{\citenamefont {Slezak}\ \emph {et~al.}(2008)\citenamefont {Slezak},
  \citenamefont {Lee}, \citenamefont {Wang}, \citenamefont {McElroy},
  \citenamefont {Fujita}, \citenamefont {Andersen}, \citenamefont {Hirschfeld},
  \citenamefont {Eisaki}, \citenamefont {Uchida},\ and\ \citenamefont
  {Davis}}]{slezak2008imaging}%
  \BibitemOpen
  \bibfield  {author} {\bibinfo {author} {\bibfnamefont {J.}~\bibnamefont
  {Slezak}}, \bibinfo {author} {\bibfnamefont {J.}~\bibnamefont {Lee}},
  \bibinfo {author} {\bibfnamefont {M.}~\bibnamefont {Wang}}, \bibinfo {author}
  {\bibfnamefont {K.}~\bibnamefont {McElroy}}, \bibinfo {author} {\bibfnamefont
  {K.}~\bibnamefont {Fujita}}, \bibinfo {author} {\bibfnamefont
  {B.}~\bibnamefont {Andersen}}, \bibinfo {author} {\bibfnamefont
  {P.}~\bibnamefont {Hirschfeld}}, \bibinfo {author} {\bibfnamefont
  {H.}~\bibnamefont {Eisaki}}, \bibinfo {author} {\bibfnamefont
  {S.}~\bibnamefont {Uchida}}, \ and\ \bibinfo {author} {\bibfnamefont
  {J.}~\bibnamefont {Davis}},\ }\href@noop {} {\bibfield  {journal} {\bibinfo
  {journal} {Proc. Natl. Acad. Sci. U. S. A.}\ }\textbf {\bibinfo {volume}
  {105}},\ \bibinfo {pages} {3203} (\bibinfo {year} {2008})}\BibitemShut
  {NoStop}%
\bibitem [{\citenamefont {Sakakibara}\ \emph {et~al.}(2012)\citenamefont
  {Sakakibara}, \citenamefont {Usui}, \citenamefont {Kuroki}, \citenamefont
  {Arita},\ and\ \citenamefont {Aoki}}]{sakakibara2012origin}%
  \BibitemOpen
  \bibfield  {author} {\bibinfo {author} {\bibfnamefont {H.}~\bibnamefont
  {Sakakibara}}, \bibinfo {author} {\bibfnamefont {H.}~\bibnamefont {Usui}},
  \bibinfo {author} {\bibfnamefont {K.}~\bibnamefont {Kuroki}}, \bibinfo
  {author} {\bibfnamefont {R.}~\bibnamefont {Arita}}, \ and\ \bibinfo {author}
  {\bibfnamefont {H.}~\bibnamefont {Aoki}},\ }\href@noop {} {\bibfield
  {journal} {\bibinfo  {journal} {Phys. Rev. B}\ }\textbf {\bibinfo {volume}
  {85}},\ \bibinfo {pages} {064501} (\bibinfo {year} {2012})}\BibitemShut
  {NoStop}%
\bibitem [{\citenamefont {Kordyuk}\ \emph {et~al.}(2005)\citenamefont
  {Kordyuk}, \citenamefont {Borisenko}, \citenamefont {Koitzsch}, \citenamefont
  {Fink}, \citenamefont {Knupfer},\ and\ \citenamefont
  {Berger}}]{kordyuk2005bare}%
  \BibitemOpen
  \bibfield  {author} {\bibinfo {author} {\bibfnamefont {A.}~\bibnamefont
  {Kordyuk}}, \bibinfo {author} {\bibfnamefont {S.}~\bibnamefont {Borisenko}},
  \bibinfo {author} {\bibfnamefont {A.}~\bibnamefont {Koitzsch}}, \bibinfo
  {author} {\bibfnamefont {J.}~\bibnamefont {Fink}}, \bibinfo {author}
  {\bibfnamefont {M.}~\bibnamefont {Knupfer}}, \ and\ \bibinfo {author}
  {\bibfnamefont {H.}~\bibnamefont {Berger}},\ }\href@noop {} {\bibfield
  {journal} {\bibinfo  {journal} {Phys. Rev. B}\ }\textbf {\bibinfo {volume}
  {71}},\ \bibinfo {pages} {214513} (\bibinfo {year} {2005})}\BibitemShut
  {NoStop}%
\bibitem [{\citenamefont {Yoshida}\ \emph {et~al.}(2006)\citenamefont
  {Yoshida}, \citenamefont {Zhou}, \citenamefont {Tanaka}, \citenamefont
  {Yang}, \citenamefont {Hussain}, \citenamefont {Shen}, \citenamefont
  {Fujimori}, \citenamefont {Sahrakorpi}, \citenamefont {Lindroos},
  \citenamefont {Markiewicz} \emph {et~al.}}]{yoshida2006systematic}%
  \BibitemOpen
  \bibfield  {author} {\bibinfo {author} {\bibfnamefont {T.}~\bibnamefont
  {Yoshida}}, \bibinfo {author} {\bibfnamefont {X.}~\bibnamefont {Zhou}},
  \bibinfo {author} {\bibfnamefont {K.}~\bibnamefont {Tanaka}}, \bibinfo
  {author} {\bibfnamefont {W.}~\bibnamefont {Yang}}, \bibinfo {author}
  {\bibfnamefont {Z.}~\bibnamefont {Hussain}}, \bibinfo {author} {\bibfnamefont
  {Z.-X.}\ \bibnamefont {Shen}}, \bibinfo {author} {\bibfnamefont
  {A.}~\bibnamefont {Fujimori}}, \bibinfo {author} {\bibfnamefont
  {S.}~\bibnamefont {Sahrakorpi}}, \bibinfo {author} {\bibfnamefont
  {M.}~\bibnamefont {Lindroos}}, \bibinfo {author} {\bibfnamefont
  {R.}~\bibnamefont {Markiewicz}},  \emph {et~al.},\ }\href@noop {} {\bibfield
  {journal} {\bibinfo  {journal} {Phys. Rev. B}\ }\textbf {\bibinfo {volume}
  {74}},\ \bibinfo {pages} {224510} (\bibinfo {year} {2006})}\BibitemShut
  {NoStop}%
\bibitem [{\citenamefont {Fournier}\ \emph {et~al.}(2010)\citenamefont
  {Fournier}, \citenamefont {Levy}, \citenamefont {Pennec}, \citenamefont
  {McChesney}, \citenamefont {Bostwick}, \citenamefont {Rotenberg},
  \citenamefont {Liang}, \citenamefont {Hardy}, \citenamefont {Bonn},
  \citenamefont {Elfimov} \emph {et~al.}}]{fournier2010loss}%
  \BibitemOpen
  \bibfield  {author} {\bibinfo {author} {\bibfnamefont {D.}~\bibnamefont
  {Fournier}}, \bibinfo {author} {\bibfnamefont {G.}~\bibnamefont {Levy}},
  \bibinfo {author} {\bibfnamefont {Y.}~\bibnamefont {Pennec}}, \bibinfo
  {author} {\bibfnamefont {J.}~\bibnamefont {McChesney}}, \bibinfo {author}
  {\bibfnamefont {A.}~\bibnamefont {Bostwick}}, \bibinfo {author}
  {\bibfnamefont {E.}~\bibnamefont {Rotenberg}}, \bibinfo {author}
  {\bibfnamefont {R.}~\bibnamefont {Liang}}, \bibinfo {author} {\bibfnamefont
  {W.}~\bibnamefont {Hardy}}, \bibinfo {author} {\bibfnamefont
  {D.}~\bibnamefont {Bonn}}, \bibinfo {author} {\bibfnamefont {I.}~\bibnamefont
  {Elfimov}},  \emph {et~al.},\ }\href@noop {} {\bibfield  {journal} {\bibinfo
  {journal} {Nat. Phys.}\ }\textbf {\bibinfo {volume} {6}},\ \bibinfo {pages}
  {905} (\bibinfo {year} {2010})}\BibitemShut {NoStop}%
\bibitem [{\citenamefont {Kohno}(2012)}]{Kohno:2012PRL}%
  \BibitemOpen
  \bibfield  {author} {\bibinfo {author} {\bibfnamefont {M.}~\bibnamefont
  {Kohno}},\ }\href@noop {} {\bibfield  {journal} {\bibinfo  {journal} {Phys.
  Rev. Lett.}\ }\textbf {\bibinfo {volume} {108}},\ \bibinfo {pages} {076401}
  (\bibinfo {year} {2012})}\BibitemShut {NoStop}%
\bibitem [{\citenamefont {Kohno}(2014)}]{PhysRevB.90.035111}%
  \BibitemOpen
  \bibfield  {author} {\bibinfo {author} {\bibfnamefont {M.}~\bibnamefont
  {Kohno}},\ }\href@noop {} {\bibfield  {journal} {\bibinfo  {journal} {Phys.
  Rev. B}\ }\textbf {\bibinfo {volume} {90}},\ \bibinfo {pages} {035111}
  (\bibinfo {year} {2014})}\BibitemShut {NoStop}%
\bibitem [{\citenamefont {Li}\ \emph {et~al.}(1991)\citenamefont {Li},
  \citenamefont {Callaway},\ and\ \citenamefont {Tan}}]{PhysRevB.44.10256}%
  \BibitemOpen
  \bibfield  {author} {\bibinfo {author} {\bibfnamefont {Q.}~\bibnamefont
  {Li}}, \bibinfo {author} {\bibfnamefont {J.}~\bibnamefont {Callaway}}, \ and\
  \bibinfo {author} {\bibfnamefont {L.}~\bibnamefont {Tan}},\ }\href@noop {}
  {\bibfield  {journal} {\bibinfo  {journal} {Phys. Rev. B}\ }\textbf {\bibinfo
  {volume} {44}},\ \bibinfo {pages} {10256} (\bibinfo {year}
  {1991})}\BibitemShut {NoStop}%
\bibitem [{\citenamefont {Dagotto}\ \emph {et~al.}(1992)\citenamefont
  {Dagotto}, \citenamefont {Ortolani},\ and\ \citenamefont
  {Scalapino}}]{Dagotto1992}%
  \BibitemOpen
  \bibfield  {author} {\bibinfo {author} {\bibfnamefont {E.}~\bibnamefont
  {Dagotto}}, \bibinfo {author} {\bibfnamefont {F.}~\bibnamefont {Ortolani}}, \
  and\ \bibinfo {author} {\bibfnamefont {D.}~\bibnamefont {Scalapino}},\
  }\href@noop {} {\bibfield  {journal} {\bibinfo  {journal} {Phys. Rev. B}\
  }\textbf {\bibinfo {volume} {46}},\ \bibinfo {pages} {3183} (\bibinfo {year}
  {1992})}\BibitemShut {NoStop}%
\bibitem [{\citenamefont {Imada}\ \emph
  {et~al.}(1998{\natexlab{a}})\citenamefont {Imada}, \citenamefont {Fujimori},\
  and\ \citenamefont {Tokura}}]{RevModPhys.70.1039}%
  \BibitemOpen
  \bibfield  {author} {\bibinfo {author} {\bibfnamefont {M.}~\bibnamefont
  {Imada}}, \bibinfo {author} {\bibfnamefont {A.}~\bibnamefont {Fujimori}}, \
  and\ \bibinfo {author} {\bibfnamefont {Y.}~\bibnamefont {Tokura}},\
  }\href@noop {} {\bibfield  {journal} {\bibinfo  {journal} {Rev. Mod. Phys.}\
  }\textbf {\bibinfo {volume} {70}},\ \bibinfo {pages} {1039} (\bibinfo {year}
  {1998}{\natexlab{a}})}\BibitemShut {NoStop}%
\bibitem [{\citenamefont {White}\ \emph {et~al.}(1989)\citenamefont {White},
  \citenamefont {Scalapino}, \citenamefont {Sugar}, \citenamefont {Loh},
  \citenamefont {Gubernatis},\ and\ \citenamefont
  {Scalettar}}]{white1989numerical}%
  \BibitemOpen
  \bibfield  {author} {\bibinfo {author} {\bibfnamefont {S.}~\bibnamefont
  {White}}, \bibinfo {author} {\bibfnamefont {D.}~\bibnamefont {Scalapino}},
  \bibinfo {author} {\bibfnamefont {R.}~\bibnamefont {Sugar}}, \bibinfo
  {author} {\bibfnamefont {E.}~\bibnamefont {Loh}}, \bibinfo {author}
  {\bibfnamefont {J.}~\bibnamefont {Gubernatis}}, \ and\ \bibinfo {author}
  {\bibfnamefont {R.}~\bibnamefont {Scalettar}},\ }\href@noop {} {\bibfield
  {journal} {\bibinfo  {journal} {Phys. Rev. B}\ }\textbf {\bibinfo {volume}
  {40}},\ \bibinfo {pages} {506} (\bibinfo {year} {1989})}\BibitemShut
  {NoStop}%
\bibitem [{\citenamefont {Foulkes}\ \emph {et~al.}(2001)\citenamefont
  {Foulkes}, \citenamefont {Mitas}, \citenamefont {Needs},\ and\ \citenamefont
  {Rajagopal}}]{foulkes2001quantum}%
  \BibitemOpen
  \bibfield  {author} {\bibinfo {author} {\bibfnamefont {W.}~\bibnamefont
  {Foulkes}}, \bibinfo {author} {\bibfnamefont {L.}~\bibnamefont {Mitas}},
  \bibinfo {author} {\bibfnamefont {R.}~\bibnamefont {Needs}}, \ and\ \bibinfo
  {author} {\bibfnamefont {G.}~\bibnamefont {Rajagopal}},\ }\href@noop {}
  {\bibfield  {journal} {\bibinfo  {journal} {Rev. Mod. Phys.}\ }\textbf
  {\bibinfo {volume} {73}},\ \bibinfo {pages} {33} (\bibinfo {year}
  {2001})}\BibitemShut {NoStop}%
\bibitem [{\citenamefont {White}(1992)}]{white1992density}%
  \BibitemOpen
  \bibfield  {author} {\bibinfo {author} {\bibfnamefont {S.~R.}\ \bibnamefont
  {White}},\ }\href@noop {} {\bibfield  {journal} {\bibinfo  {journal} {Phys.
  Rev. Lett.}\ }\textbf {\bibinfo {volume} {69}},\ \bibinfo {pages} {2863}
  (\bibinfo {year} {1992})}\BibitemShut {NoStop}%
\bibitem [{\citenamefont {Schollw{\"o}ck}(2005)}]{schollwock2005density}%
  \BibitemOpen
  \bibfield  {author} {\bibinfo {author} {\bibfnamefont {U.}~\bibnamefont
  {Schollw{\"o}ck}},\ }\href@noop {} {\bibfield  {journal} {\bibinfo  {journal}
  {Rev. Mod. Phys.}\ }\textbf {\bibinfo {volume} {77}},\ \bibinfo {pages} {259}
  (\bibinfo {year} {2005})}\BibitemShut {NoStop}%
\bibitem [{\citenamefont {Kancharla}\ \emph {et~al.}(2008)\citenamefont
  {Kancharla}, \citenamefont {Kyung}, \citenamefont {S{\'e}n{\'e}chal},
  \citenamefont {Civelli}, \citenamefont {Capone}, \citenamefont {Kotliar},\
  and\ \citenamefont {Tremblay}}]{kancharla2008anomalous}%
  \BibitemOpen
  \bibfield  {author} {\bibinfo {author} {\bibfnamefont {S.}~\bibnamefont
  {Kancharla}}, \bibinfo {author} {\bibfnamefont {B.}~\bibnamefont {Kyung}},
  \bibinfo {author} {\bibfnamefont {D.}~\bibnamefont {S{\'e}n{\'e}chal}},
  \bibinfo {author} {\bibfnamefont {M.}~\bibnamefont {Civelli}}, \bibinfo
  {author} {\bibfnamefont {M.}~\bibnamefont {Capone}}, \bibinfo {author}
  {\bibfnamefont {G.}~\bibnamefont {Kotliar}}, \ and\ \bibinfo {author}
  {\bibfnamefont {A.-M.}\ \bibnamefont {Tremblay}},\ }\href@noop {} {\bibfield
  {journal} {\bibinfo  {journal} {Phys. Rev. B}\ }\textbf {\bibinfo {volume}
  {77}},\ \bibinfo {pages} {184516} (\bibinfo {year} {2008})}\BibitemShut
  {NoStop}%
\bibitem [{\citenamefont {Weber}\ \emph {et~al.}(2010)\citenamefont {Weber},
  \citenamefont {Haule},\ and\ \citenamefont {Kotliar}}]{weber2010strength}%
  \BibitemOpen
  \bibfield  {author} {\bibinfo {author} {\bibfnamefont {C.}~\bibnamefont
  {Weber}}, \bibinfo {author} {\bibfnamefont {K.}~\bibnamefont {Haule}}, \ and\
  \bibinfo {author} {\bibfnamefont {G.}~\bibnamefont {Kotliar}},\ }\href@noop
  {} {\bibfield  {journal} {\bibinfo  {journal} {Nat. Phys.}\ }\textbf
  {\bibinfo {volume} {6}},\ \bibinfo {pages} {574} (\bibinfo {year}
  {2010})}\BibitemShut {NoStop}%
\bibitem [{\citenamefont {Gull}\ \emph {et~al.}(2013)\citenamefont {Gull},
  \citenamefont {Parcollet},\ and\ \citenamefont
  {Millis}}]{gull2013superconductivity}%
  \BibitemOpen
  \bibfield  {author} {\bibinfo {author} {\bibfnamefont {E.}~\bibnamefont
  {Gull}}, \bibinfo {author} {\bibfnamefont {O.}~\bibnamefont {Parcollet}}, \
  and\ \bibinfo {author} {\bibfnamefont {A.~J.}\ \bibnamefont {Millis}},\
  }\href@noop {} {\bibfield  {journal} {\bibinfo  {journal} {Phys. Rev. Lett.}\
  }\textbf {\bibinfo {volume} {110}},\ \bibinfo {pages} {216405} (\bibinfo
  {year} {2013})}\BibitemShut {NoStop}%
\bibitem [{\citenamefont {Sakai}\ \emph {et~al.}(2009)\citenamefont {Sakai},
  \citenamefont {Motome},\ and\ \citenamefont {Imada}}]{sakai2009evolution}%
  \BibitemOpen
  \bibfield  {author} {\bibinfo {author} {\bibfnamefont {S.}~\bibnamefont
  {Sakai}}, \bibinfo {author} {\bibfnamefont {Y.}~\bibnamefont {Motome}}, \
  and\ \bibinfo {author} {\bibfnamefont {M.}~\bibnamefont {Imada}},\
  }\href@noop {} {\bibfield  {journal} {\bibinfo  {journal} {Phys. Rev. Lett.}\
  }\textbf {\bibinfo {volume} {102}},\ \bibinfo {pages} {056404} (\bibinfo
  {year} {2009})}\BibitemShut {NoStop}%
\bibitem [{\citenamefont {Gros}\ and\ \citenamefont
  {Valenti}(1993)}]{gros1993cluster}%
  \BibitemOpen
  \bibfield  {author} {\bibinfo {author} {\bibfnamefont {C.}~\bibnamefont
  {Gros}}\ and\ \bibinfo {author} {\bibfnamefont {R.}~\bibnamefont {Valenti}},\
  }\href@noop {} {\bibfield  {journal} {\bibinfo  {journal} {Phys. Rev. B}\
  }\textbf {\bibinfo {volume} {48}},\ \bibinfo {pages} {418} (\bibinfo {year}
  {1993})}\BibitemShut {NoStop}%
\bibitem [{\citenamefont {S{\'e}n{\'e}chal}\ \emph {et~al.}(2000)\citenamefont
  {S{\'e}n{\'e}chal}, \citenamefont {Perez},\ and\ \citenamefont
  {Pioro-Ladri{\`e}re}}]{Senechal:2000fg}%
  \BibitemOpen
  \bibfield  {author} {\bibinfo {author} {\bibfnamefont {D.}~\bibnamefont
  {S{\'e}n{\'e}chal}}, \bibinfo {author} {\bibfnamefont {D.}~\bibnamefont
  {Perez}}, \ and\ \bibinfo {author} {\bibfnamefont {M.}~\bibnamefont
  {Pioro-Ladri{\`e}re}},\ }\href@noop {} {\bibfield  {journal} {\bibinfo
  {journal} {Phys. Rev. Lett.}\ }\textbf {\bibinfo {volume} {84}},\ \bibinfo
  {pages} {522} (\bibinfo {year} {2000})}\BibitemShut {NoStop}%
\bibitem [{\citenamefont {S{\'e}n{\'e}chal}\ \emph {et~al.}(2002)\citenamefont
  {S{\'e}n{\'e}chal}, \citenamefont {Perez},\ and\ \citenamefont
  {Plouffe}}]{Senechal:2002fr}%
  \BibitemOpen
  \bibfield  {author} {\bibinfo {author} {\bibfnamefont {D.}~\bibnamefont
  {S{\'e}n{\'e}chal}}, \bibinfo {author} {\bibfnamefont {D.}~\bibnamefont
  {Perez}}, \ and\ \bibinfo {author} {\bibfnamefont {D.}~\bibnamefont
  {Plouffe}},\ }\href@noop {} {\bibfield  {journal} {\bibinfo  {journal} {Phys.
  Rev. B}\ }\textbf {\bibinfo {volume} {66}},\ \bibinfo {pages} {075129}
  (\bibinfo {year} {2002})}\BibitemShut {NoStop}%
\bibitem [{\citenamefont {Yuan}\ \emph {et~al.}(2005)\citenamefont {Yuan},
  \citenamefont {Yuan},\ and\ \citenamefont {Ting}}]{yuan2005doping}%
  \BibitemOpen
  \bibfield  {author} {\bibinfo {author} {\bibfnamefont {Q.}~\bibnamefont
  {Yuan}}, \bibinfo {author} {\bibfnamefont {F.}~\bibnamefont {Yuan}}, \ and\
  \bibinfo {author} {\bibfnamefont {C.}~\bibnamefont {Ting}},\ }\href@noop {}
  {\bibfield  {journal} {\bibinfo  {journal} {Phys. Rev. B}\ }\textbf {\bibinfo
  {volume} {72}},\ \bibinfo {pages} {054504} (\bibinfo {year}
  {2005})}\BibitemShut {NoStop}%
\bibitem [{\citenamefont {Arrigoni}\ \emph {et~al.}(2009)\citenamefont
  {Arrigoni}, \citenamefont {Aichhorn}, \citenamefont {Daghofer},\ and\
  \citenamefont {Hanke}}]{arrigoni2009phase}%
  \BibitemOpen
  \bibfield  {author} {\bibinfo {author} {\bibfnamefont {E.}~\bibnamefont
  {Arrigoni}}, \bibinfo {author} {\bibfnamefont {M.}~\bibnamefont {Aichhorn}},
  \bibinfo {author} {\bibfnamefont {M.}~\bibnamefont {Daghofer}}, \ and\
  \bibinfo {author} {\bibfnamefont {W.}~\bibnamefont {Hanke}},\ }\href@noop {}
  {\bibfield  {journal} {\bibinfo  {journal} {New Journal of Physics}\ }\textbf
  {\bibinfo {volume} {11}},\ \bibinfo {pages} {055066} (\bibinfo {year}
  {2009})}\BibitemShut {NoStop}%
\bibitem [{\citenamefont {Han}\ \emph {et~al.}(2016)\citenamefont {Han},
  \citenamefont {Liu}, \citenamefont {Liu}, \citenamefont {Li}, \citenamefont
  {Chen}, \citenamefont {Liao}, \citenamefont {Xie}, \citenamefont {Normand},\
  and\ \citenamefont {Xiang}}]{han2016charge}%
  \BibitemOpen
  \bibfield  {author} {\bibinfo {author} {\bibfnamefont {X.-J.}\ \bibnamefont
  {Han}}, \bibinfo {author} {\bibfnamefont {Y.}~\bibnamefont {Liu}}, \bibinfo
  {author} {\bibfnamefont {Z.-Y.}\ \bibnamefont {Liu}}, \bibinfo {author}
  {\bibfnamefont {X.}~\bibnamefont {Li}}, \bibinfo {author} {\bibfnamefont
  {J.}~\bibnamefont {Chen}}, \bibinfo {author} {\bibfnamefont {H.-J.}\
  \bibnamefont {Liao}}, \bibinfo {author} {\bibfnamefont {Z.-Y.}\ \bibnamefont
  {Xie}}, \bibinfo {author} {\bibfnamefont {B.}~\bibnamefont {Normand}}, \ and\
  \bibinfo {author} {\bibfnamefont {T.}~\bibnamefont {Xiang}},\ }\href@noop {}
  {\bibfield  {journal} {\bibinfo  {journal} {New J Phys.}\ }\textbf {\bibinfo
  {volume} {18}},\ \bibinfo {pages} {103004} (\bibinfo {year}
  {2016})}\BibitemShut {NoStop}%
\bibitem [{\citenamefont {Hettler}\ \emph {et~al.}(2000)\citenamefont
  {Hettler}, \citenamefont {Mukherjee}, \citenamefont {Jarrell},\ and\
  \citenamefont {Krishnamurthy}}]{hettler2000dynamical}%
  \BibitemOpen
  \bibfield  {author} {\bibinfo {author} {\bibfnamefont {M.}~\bibnamefont
  {Hettler}}, \bibinfo {author} {\bibfnamefont {M.}~\bibnamefont {Mukherjee}},
  \bibinfo {author} {\bibfnamefont {M.}~\bibnamefont {Jarrell}}, \ and\
  \bibinfo {author} {\bibfnamefont {H.}~\bibnamefont {Krishnamurthy}},\
  }\href@noop {} {\bibfield  {journal} {\bibinfo  {journal} {Phys. Rev. B}\
  }\textbf {\bibinfo {volume} {61}},\ \bibinfo {pages} {12739} (\bibinfo {year}
  {2000})}\BibitemShut {NoStop}%
\bibitem [{\citenamefont {Potthoff}\ \emph {et~al.}(2003)\citenamefont
  {Potthoff}, \citenamefont {Aichhorn},\ and\ \citenamefont
  {Dahnken}}]{potthoff2003variational}%
  \BibitemOpen
  \bibfield  {author} {\bibinfo {author} {\bibfnamefont {M.}~\bibnamefont
  {Potthoff}}, \bibinfo {author} {\bibfnamefont {M.}~\bibnamefont {Aichhorn}},
  \ and\ \bibinfo {author} {\bibfnamefont {C.}~\bibnamefont {Dahnken}},\
  }\href@noop {} {\bibfield  {journal} {\bibinfo  {journal} {Phys. Rev. Lett.}\
  }\textbf {\bibinfo {volume} {91}},\ \bibinfo {pages} {206402} (\bibinfo
  {year} {2003})}\BibitemShut {NoStop}%
\bibitem [{\citenamefont {Wang}\ \emph {et~al.}(2015)\citenamefont {Wang},
  \citenamefont {Wohlfeld}, \citenamefont {Moritz}, \citenamefont {Jia},
  \citenamefont {van Veenendaal}, \citenamefont {Wu}, \citenamefont {Chen},\
  and\ \citenamefont {Devereaux}}]{wang2015origin}%
  \BibitemOpen
  \bibfield  {author} {\bibinfo {author} {\bibfnamefont {Y.}~\bibnamefont
  {Wang}}, \bibinfo {author} {\bibfnamefont {K.}~\bibnamefont {Wohlfeld}},
  \bibinfo {author} {\bibfnamefont {B.}~\bibnamefont {Moritz}}, \bibinfo
  {author} {\bibfnamefont {C.}~\bibnamefont {Jia}}, \bibinfo {author}
  {\bibfnamefont {M.}~\bibnamefont {van Veenendaal}}, \bibinfo {author}
  {\bibfnamefont {K.}~\bibnamefont {Wu}}, \bibinfo {author} {\bibfnamefont
  {C.-C.}\ \bibnamefont {Chen}}, \ and\ \bibinfo {author} {\bibfnamefont
  {T.~P.}\ \bibnamefont {Devereaux}},\ }\href@noop {} {\bibfield  {journal}
  {\bibinfo  {journal} {Phys. Rev. B}\ }\textbf {\bibinfo {volume} {92}},\
  \bibinfo {pages} {075119} (\bibinfo {year} {2015})}\BibitemShut {NoStop}%
\bibitem [{\citenamefont {Wang}\ \emph
  {et~al.}(2018{\natexlab{a}})\citenamefont {Wang}, \citenamefont {Wohlfeld},
  \citenamefont {Moritz}, \citenamefont {Jia}, \citenamefont {van Veenendaal},
  \citenamefont {Wu}, \citenamefont {Chen},\ and\ \citenamefont
  {Devereaux}}]{wang2018erratum}%
  \BibitemOpen
  \bibfield  {author} {\bibinfo {author} {\bibfnamefont {Y.}~\bibnamefont
  {Wang}}, \bibinfo {author} {\bibfnamefont {K.}~\bibnamefont {Wohlfeld}},
  \bibinfo {author} {\bibfnamefont {B.}~\bibnamefont {Moritz}}, \bibinfo
  {author} {\bibfnamefont {C.~J.}\ \bibnamefont {Jia}}, \bibinfo {author}
  {\bibfnamefont {M.}~\bibnamefont {van Veenendaal}}, \bibinfo {author}
  {\bibfnamefont {K.}~\bibnamefont {Wu}}, \bibinfo {author} {\bibfnamefont
  {C.-C.}\ \bibnamefont {Chen}}, \ and\ \bibinfo {author} {\bibfnamefont
  {T.~P.}\ \bibnamefont {Devereaux}},\ }\href@noop {} {\bibfield  {journal}
  {\bibinfo  {journal} {Phys. Rev. B}\ }\textbf {\bibinfo {volume} {97}},\
  \bibinfo {pages} {199903} (\bibinfo {year} {2018}{\natexlab{a}})}\BibitemShut
  {NoStop}%
\bibitem [{\citenamefont {Moritz}\ \emph {et~al.}(2009)\citenamefont {Moritz},
  \citenamefont {Schmitt}, \citenamefont {Meevasana}, \citenamefont {Johnston},
  \citenamefont {Motoyama}, \citenamefont {Greven}, \citenamefont {Lu},
  \citenamefont {Kim}, \citenamefont {Scalettar}, \citenamefont {Shen},\ and\
  \citenamefont {Devereaux}}]{moritz2009effect}%
  \BibitemOpen
  \bibfield  {author} {\bibinfo {author} {\bibfnamefont {B.}~\bibnamefont
  {Moritz}}, \bibinfo {author} {\bibfnamefont {F.}~\bibnamefont {Schmitt}},
  \bibinfo {author} {\bibfnamefont {W.}~\bibnamefont {Meevasana}}, \bibinfo
  {author} {\bibfnamefont {S.}~\bibnamefont {Johnston}}, \bibinfo {author}
  {\bibfnamefont {E.~M.}\ \bibnamefont {Motoyama}}, \bibinfo {author}
  {\bibfnamefont {M.}~\bibnamefont {Greven}}, \bibinfo {author} {\bibfnamefont
  {D.~H.}\ \bibnamefont {Lu}}, \bibinfo {author} {\bibfnamefont
  {C.}~\bibnamefont {Kim}}, \bibinfo {author} {\bibfnamefont {R.~T.}\
  \bibnamefont {Scalettar}}, \bibinfo {author} {\bibfnamefont {Z.~X.}\
  \bibnamefont {Shen}}, \ and\ \bibinfo {author} {\bibfnamefont {T.~P.}\
  \bibnamefont {Devereaux}},\ }\href@noop {} {\bibfield  {journal} {\bibinfo
  {journal} {New J Phys.}\ }\textbf {\bibinfo {volume} {11}},\ \bibinfo {pages}
  {093020} (\bibinfo {year} {2009})}\BibitemShut {NoStop}%
\bibitem [{\citenamefont {Wang}\ \emph
  {et~al.}(2018{\natexlab{b}})\citenamefont {Wang}, \citenamefont {Moritz},
  \citenamefont {Chen}, \citenamefont {Devereaux},\ and\ \citenamefont
  {Wohlfeld}}]{wang2018influence}%
  \BibitemOpen
  \bibfield  {author} {\bibinfo {author} {\bibfnamefont {Y.}~\bibnamefont
  {Wang}}, \bibinfo {author} {\bibfnamefont {B.}~\bibnamefont {Moritz}},
  \bibinfo {author} {\bibfnamefont {C.-C.}\ \bibnamefont {Chen}}, \bibinfo
  {author} {\bibfnamefont {T.~P.}\ \bibnamefont {Devereaux}}, \ and\ \bibinfo
  {author} {\bibfnamefont {K.}~\bibnamefont {Wohlfeld}},\ }\href@noop {}
  {\bibfield  {journal} {\bibinfo  {journal} {Phys. Rev. B}\ }\textbf {\bibinfo
  {volume} {97}},\ \bibinfo {pages} {115120} (\bibinfo {year}
  {2018}{\natexlab{b}})}\BibitemShut {NoStop}%
\bibitem [{\citenamefont {Zhong}\ \emph {et~al.}(2020)\citenamefont {Zhong},
  \citenamefont {Fan}, \citenamefont {Wang}, \citenamefont {Wang},
  \citenamefont {Zhang}, \citenamefont {Zhu}, \citenamefont {Dou},
  \citenamefont {Yu}, \citenamefont {Wang}, \citenamefont {Zhang} \emph
  {et~al.}}]{zhong2020direct}%
  \BibitemOpen
  \bibfield  {author} {\bibinfo {author} {\bibfnamefont {Y.}~\bibnamefont
  {Zhong}}, \bibinfo {author} {\bibfnamefont {J.-Q.}\ \bibnamefont {Fan}},
  \bibinfo {author} {\bibfnamefont {R.-F.}\ \bibnamefont {Wang}}, \bibinfo
  {author} {\bibfnamefont {S.}~\bibnamefont {Wang}}, \bibinfo {author}
  {\bibfnamefont {X.}~\bibnamefont {Zhang}}, \bibinfo {author} {\bibfnamefont
  {Y.}~\bibnamefont {Zhu}}, \bibinfo {author} {\bibfnamefont {Z.}~\bibnamefont
  {Dou}}, \bibinfo {author} {\bibfnamefont {X.-Q.}\ \bibnamefont {Yu}},
  \bibinfo {author} {\bibfnamefont {Y.}~\bibnamefont {Wang}}, \bibinfo {author}
  {\bibfnamefont {D.}~\bibnamefont {Zhang}},  \emph {et~al.},\ }\href@noop {}
  {\bibfield  {journal} {\bibinfo  {journal} {Phys. Rev. Lett.}\ }\textbf
  {\bibinfo {volume} {125}},\ \bibinfo {pages} {077002} (\bibinfo {year}
  {2020})}\BibitemShut {NoStop}%
\bibitem [{\citenamefont {Imada}\ \emph
  {et~al.}(1998{\natexlab{b}})\citenamefont {Imada}, \citenamefont {Fujimori},\
  and\ \citenamefont {Tokura}}]{imada1998metal}%
  \BibitemOpen
  \bibfield  {author} {\bibinfo {author} {\bibfnamefont {M.}~\bibnamefont
  {Imada}}, \bibinfo {author} {\bibfnamefont {A.}~\bibnamefont {Fujimori}}, \
  and\ \bibinfo {author} {\bibfnamefont {Y.}~\bibnamefont {Tokura}},\
  }\href@noop {} {\bibfield  {journal} {\bibinfo  {journal} {Rev. Mod. Phys.}\
  }\textbf {\bibinfo {volume} {70}},\ \bibinfo {pages} {1039} (\bibinfo {year}
  {1998}{\natexlab{b}})}\BibitemShut {NoStop}%
\bibitem [{\citenamefont {Eskes}\ \emph {et~al.}(1991)\citenamefont {Eskes},
  \citenamefont {Meinders},\ and\ \citenamefont
  {Sawatzky}}]{eskes1991anomalous}%
  \BibitemOpen
  \bibfield  {author} {\bibinfo {author} {\bibfnamefont {H.}~\bibnamefont
  {Eskes}}, \bibinfo {author} {\bibfnamefont {M.}~\bibnamefont {Meinders}}, \
  and\ \bibinfo {author} {\bibfnamefont {G.}~\bibnamefont {Sawatzky}},\
  }\href@noop {} {\bibfield  {journal} {\bibinfo  {journal} {Phys. Rev. Lett.}\
  }\textbf {\bibinfo {volume} {67}},\ \bibinfo {pages} {1035} (\bibinfo {year}
  {1991})}\BibitemShut {NoStop}%
\bibitem [{\citenamefont {Dagotto}\ \emph {et~al.}(1991)\citenamefont
  {Dagotto}, \citenamefont {Moreo}, \citenamefont {Ortolani}, \citenamefont
  {Riera},\ and\ \citenamefont {Scalapino}}]{dagotto1991density}%
  \BibitemOpen
  \bibfield  {author} {\bibinfo {author} {\bibfnamefont {E.}~\bibnamefont
  {Dagotto}}, \bibinfo {author} {\bibfnamefont {A.}~\bibnamefont {Moreo}},
  \bibinfo {author} {\bibfnamefont {F.}~\bibnamefont {Ortolani}}, \bibinfo
  {author} {\bibfnamefont {J.}~\bibnamefont {Riera}}, \ and\ \bibinfo {author}
  {\bibfnamefont {D.}~\bibnamefont {Scalapino}},\ }\href@noop {} {\bibfield
  {journal} {\bibinfo  {journal} {Phys. Rev. Lett.}\ }\textbf {\bibinfo
  {volume} {67}},\ \bibinfo {pages} {1918} (\bibinfo {year}
  {1991})}\BibitemShut {NoStop}%
\bibitem [{\citenamefont {Chen}\ \emph {et~al.}(1991)\citenamefont {Chen},
  \citenamefont {Sette}, \citenamefont {Ma}, \citenamefont {Hybertsen},
  \citenamefont {Stechel}, \citenamefont {Foulkes}, \citenamefont {Schulter},
  \citenamefont {Cheong}, \citenamefont {Cooper}, \citenamefont {Rupp~Jr} \emph
  {et~al.}}]{chen1991electronic}%
  \BibitemOpen
  \bibfield  {author} {\bibinfo {author} {\bibfnamefont {C.}~\bibnamefont
  {Chen}}, \bibinfo {author} {\bibfnamefont {F.}~\bibnamefont {Sette}},
  \bibinfo {author} {\bibfnamefont {Y.}~\bibnamefont {Ma}}, \bibinfo {author}
  {\bibfnamefont {M.}~\bibnamefont {Hybertsen}}, \bibinfo {author}
  {\bibfnamefont {E.}~\bibnamefont {Stechel}}, \bibinfo {author} {\bibfnamefont
  {W.}~\bibnamefont {Foulkes}}, \bibinfo {author} {\bibfnamefont
  {M.}~\bibnamefont {Schulter}}, \bibinfo {author} {\bibfnamefont
  {S.}~\bibnamefont {Cheong}}, \bibinfo {author} {\bibfnamefont
  {A.}~\bibnamefont {Cooper}}, \bibinfo {author} {\bibfnamefont
  {L.}~\bibnamefont {Rupp~Jr}},  \emph {et~al.},\ }\href@noop {} {\bibfield
  {journal} {\bibinfo  {journal} {Phys. Rev. Lett.}\ }\textbf {\bibinfo
  {volume} {66}},\ \bibinfo {pages} {104} (\bibinfo {year} {1991})}\BibitemShut
  {NoStop}%
\bibitem [{\citenamefont {Liebsch}(2010)}]{liebsch2010spectral}%
  \BibitemOpen
  \bibfield  {author} {\bibinfo {author} {\bibfnamefont {A.}~\bibnamefont
  {Liebsch}},\ }\href@noop {} {\bibfield  {journal} {\bibinfo  {journal} {Phys.
  Rev. B}\ }\textbf {\bibinfo {volume} {81}},\ \bibinfo {pages} {235133}
  (\bibinfo {year} {2010})}\BibitemShut {NoStop}%
\bibitem [{\citenamefont {Phillips}(2010)}]{phillips2010colloquium}%
  \BibitemOpen
  \bibfield  {author} {\bibinfo {author} {\bibfnamefont {P.}~\bibnamefont
  {Phillips}},\ }\href@noop {} {\bibfield  {journal} {\bibinfo  {journal} {Rev.
  Mod. Phys.}\ }\textbf {\bibinfo {volume} {82}},\ \bibinfo {pages} {1719}
  (\bibinfo {year} {2010})}\BibitemShut {NoStop}%
\bibitem [{\citenamefont {Armitage}\ \emph {et~al.}(2010)\citenamefont
  {Armitage}, \citenamefont {Fournier},\ and\ \citenamefont
  {Greene}}]{armitage2010progress}%
  \BibitemOpen
  \bibfield  {author} {\bibinfo {author} {\bibfnamefont {N.}~\bibnamefont
  {Armitage}}, \bibinfo {author} {\bibfnamefont {P.}~\bibnamefont {Fournier}},
  \ and\ \bibinfo {author} {\bibfnamefont {R.}~\bibnamefont {Greene}},\
  }\href@noop {} {\bibfield  {journal} {\bibinfo  {journal} {Rev. Mod. Phys.}\
  }\textbf {\bibinfo {volume} {82}},\ \bibinfo {pages} {2421} (\bibinfo {year}
  {2010})}\BibitemShut {NoStop}%
\bibitem [{\citenamefont {S{\'e}n{\'e}chal}\ and\ \citenamefont
  {Tremblay}(2004)}]{senechal2004hot}%
  \BibitemOpen
  \bibfield  {author} {\bibinfo {author} {\bibfnamefont {D.}~\bibnamefont
  {S{\'e}n{\'e}chal}}\ and\ \bibinfo {author} {\bibfnamefont {A.-M.}\
  \bibnamefont {Tremblay}},\ }\href@noop {} {\bibfield  {journal} {\bibinfo
  {journal} {Phys. Rev. Lett.}\ }\textbf {\bibinfo {volume} {92}},\ \bibinfo
  {pages} {126401} (\bibinfo {year} {2004})}\BibitemShut {NoStop}%
\bibitem [{\citenamefont {Yagi}\ \emph {et~al.}(2006)\citenamefont {Yagi},
  \citenamefont {Yoshida}, \citenamefont {Fujimori}, \citenamefont {Kohsaka},
  \citenamefont {Misawa}, \citenamefont {Sasagawa}, \citenamefont {Takagi},
  \citenamefont {Azuma},\ and\ \citenamefont {Takano}}]{yagi2006chemical}%
  \BibitemOpen
  \bibfield  {author} {\bibinfo {author} {\bibfnamefont {H.}~\bibnamefont
  {Yagi}}, \bibinfo {author} {\bibfnamefont {T.}~\bibnamefont {Yoshida}},
  \bibinfo {author} {\bibfnamefont {A.}~\bibnamefont {Fujimori}}, \bibinfo
  {author} {\bibfnamefont {Y.}~\bibnamefont {Kohsaka}}, \bibinfo {author}
  {\bibfnamefont {M.}~\bibnamefont {Misawa}}, \bibinfo {author} {\bibfnamefont
  {T.}~\bibnamefont {Sasagawa}}, \bibinfo {author} {\bibfnamefont
  {H.}~\bibnamefont {Takagi}}, \bibinfo {author} {\bibfnamefont
  {M.}~\bibnamefont {Azuma}}, \ and\ \bibinfo {author} {\bibfnamefont
  {M.}~\bibnamefont {Takano}},\ }\href@noop {} {\bibfield  {journal} {\bibinfo
  {journal} {Phys. Rev. B}\ }\textbf {\bibinfo {volume} {73}},\ \bibinfo
  {pages} {172503} (\bibinfo {year} {2006})}\BibitemShut {NoStop}%
\bibitem [{\citenamefont {Tohyama}(2004)}]{tohyama2004asymmetry}%
  \BibitemOpen
  \bibfield  {author} {\bibinfo {author} {\bibfnamefont {T.}~\bibnamefont
  {Tohyama}},\ }\href@noop {} {\bibfield  {journal} {\bibinfo  {journal} {Phys.
  Rev. B}\ }\textbf {\bibinfo {volume} {70}},\ \bibinfo {pages} {174517}
  (\bibinfo {year} {2004})}\BibitemShut {NoStop}%
\bibitem [{\citenamefont {Moritz}\ \emph {et~al.}(2011)\citenamefont {Moritz},
  \citenamefont {Johnston}, \citenamefont {Devereaux}, \citenamefont
  {Muschler}, \citenamefont {Prestel}, \citenamefont {Hackl}, \citenamefont
  {Lambacher}, \citenamefont {Erb}, \citenamefont {Komiya},\ and\ \citenamefont
  {Ando}}]{moritz2011investigation}%
  \BibitemOpen
  \bibfield  {author} {\bibinfo {author} {\bibfnamefont {B.}~\bibnamefont
  {Moritz}}, \bibinfo {author} {\bibfnamefont {S.}~\bibnamefont {Johnston}},
  \bibinfo {author} {\bibfnamefont {T.}~\bibnamefont {Devereaux}}, \bibinfo
  {author} {\bibfnamefont {B.}~\bibnamefont {Muschler}}, \bibinfo {author}
  {\bibfnamefont {W.}~\bibnamefont {Prestel}}, \bibinfo {author} {\bibfnamefont
  {R.}~\bibnamefont {Hackl}}, \bibinfo {author} {\bibfnamefont
  {M.}~\bibnamefont {Lambacher}}, \bibinfo {author} {\bibfnamefont
  {A.}~\bibnamefont {Erb}}, \bibinfo {author} {\bibfnamefont {S.}~\bibnamefont
  {Komiya}}, \ and\ \bibinfo {author} {\bibfnamefont {Y.}~\bibnamefont
  {Ando}},\ }\href@noop {} {\bibfield  {journal} {\bibinfo  {journal} {Phys.
  Rev. B}\ }\textbf {\bibinfo {volume} {84}},\ \bibinfo {pages} {235114}
  (\bibinfo {year} {2011})}\BibitemShut {NoStop}%
\bibitem [{\citenamefont {Wang}\ \emph {et~al.}(2014)\citenamefont {Wang},
  \citenamefont {Jia}, \citenamefont {Moritz},\ and\ \citenamefont
  {Devereaux}}]{wang2014real}%
  \BibitemOpen
  \bibfield  {author} {\bibinfo {author} {\bibfnamefont {Y.}~\bibnamefont
  {Wang}}, \bibinfo {author} {\bibfnamefont {C.}~\bibnamefont {Jia}}, \bibinfo
  {author} {\bibfnamefont {B.}~\bibnamefont {Moritz}}, \ and\ \bibinfo {author}
  {\bibfnamefont {T.~P.}\ \bibnamefont {Devereaux}},\ }\href@noop {} {\bibfield
   {journal} {\bibinfo  {journal} {Phys. Rev. Lett.}\ }\textbf {\bibinfo
  {volume} {112}},\ \bibinfo {pages} {156402} (\bibinfo {year}
  {2014})}\BibitemShut {NoStop}%
\bibitem [{\citenamefont {Vishik}\ \emph {et~al.}(2014)\citenamefont {Vishik},
  \citenamefont {Bari{\v{s}}i{\'c}}, \citenamefont {Chan}, \citenamefont {Li},
  \citenamefont {Xia}, \citenamefont {Yu}, \citenamefont {Zhao}, \citenamefont
  {Lee}, \citenamefont {Meevasana}, \citenamefont {Devereaux} \emph
  {et~al.}}]{vishik2014angle}%
  \BibitemOpen
  \bibfield  {author} {\bibinfo {author} {\bibfnamefont {I.}~\bibnamefont
  {Vishik}}, \bibinfo {author} {\bibfnamefont {N.}~\bibnamefont
  {Bari{\v{s}}i{\'c}}}, \bibinfo {author} {\bibfnamefont {M.}~\bibnamefont
  {Chan}}, \bibinfo {author} {\bibfnamefont {Y.}~\bibnamefont {Li}}, \bibinfo
  {author} {\bibfnamefont {D.}~\bibnamefont {Xia}}, \bibinfo {author}
  {\bibfnamefont {G.}~\bibnamefont {Yu}}, \bibinfo {author} {\bibfnamefont
  {X.}~\bibnamefont {Zhao}}, \bibinfo {author} {\bibfnamefont {W.}~\bibnamefont
  {Lee}}, \bibinfo {author} {\bibfnamefont {W.}~\bibnamefont {Meevasana}},
  \bibinfo {author} {\bibfnamefont {T.}~\bibnamefont {Devereaux}},  \emph
  {et~al.},\ }\href@noop {} {\bibfield  {journal} {\bibinfo  {journal} {Phys.
  Rev. B}\ }\textbf {\bibinfo {volume} {89}},\ \bibinfo {pages} {195141}
  (\bibinfo {year} {2014})}\BibitemShut {NoStop}%
\bibitem [{\citenamefont {Zhang}\ and\ \citenamefont
  {Rice}(1988{\natexlab{b}})}]{Zhang:1988jf}%
  \BibitemOpen
  \bibfield  {author} {\bibinfo {author} {\bibfnamefont {F.}~\bibnamefont
  {Zhang}}\ and\ \bibinfo {author} {\bibfnamefont {T.}~\bibnamefont {Rice}},\
  }\href@noop {} {\bibfield  {journal} {\bibinfo  {journal} {Phys.Rev. B}\
  }\textbf {\bibinfo {volume} {37}},\ \bibinfo {pages} {3759} (\bibinfo {year}
  {1988}{\natexlab{b}})}\BibitemShut {NoStop}%
\bibitem [{\citenamefont {Eskes}\ and\ \citenamefont
  {Sawatzky}(1988)}]{Eskes:1988ef}%
  \BibitemOpen
  \bibfield  {author} {\bibinfo {author} {\bibfnamefont {H.}~\bibnamefont
  {Eskes}}\ and\ \bibinfo {author} {\bibfnamefont {G.}~\bibnamefont
  {Sawatzky}},\ }\href@noop {} {\bibfield  {journal} {\bibinfo  {journal}
  {Phys. Rev. Lett.}\ }\textbf {\bibinfo {volume} {61}},\ \bibinfo {pages}
  {1415} (\bibinfo {year} {1988})}\BibitemShut {NoStop}%
\bibitem [{\citenamefont {Lehoucq}\ \emph {et~al.}(1998)\citenamefont
  {Lehoucq}, \citenamefont {Sorensen},\ and\ \citenamefont
  {Yang}}]{lehoucq1998arpack}%
  \BibitemOpen
  \bibfield  {author} {\bibinfo {author} {\bibfnamefont {R.~B.}\ \bibnamefont
  {Lehoucq}}, \bibinfo {author} {\bibfnamefont {D.~C.}\ \bibnamefont
  {Sorensen}}, \ and\ \bibinfo {author} {\bibfnamefont {C.}~\bibnamefont
  {Yang}},\ }\href@noop {} {\emph {\bibinfo {title} {ARPACK Users' Guide:
  Solution of Large-Scale Eigenvalue Problems with Implicitly Restarted Arnoldi
  Methods}}}\ (\bibinfo  {publisher} {Siam},\ \bibinfo {year}
  {1998})\BibitemShut {NoStop}%
\bibitem [{\citenamefont {Jia}\ \emph {et~al.}(2018)\citenamefont {Jia},
  \citenamefont {Wang}, \citenamefont {Mendl}, \citenamefont {Moritz},\ and\
  \citenamefont {Devereaux}}]{jia2017paradeisos}%
  \BibitemOpen
  \bibfield  {author} {\bibinfo {author} {\bibfnamefont {C.}~\bibnamefont
  {Jia}}, \bibinfo {author} {\bibfnamefont {Y.}~\bibnamefont {Wang}}, \bibinfo
  {author} {\bibfnamefont {C.}~\bibnamefont {Mendl}}, \bibinfo {author}
  {\bibfnamefont {B.}~\bibnamefont {Moritz}}, \ and\ \bibinfo {author}
  {\bibfnamefont {T.}~\bibnamefont {Devereaux}},\ }\href@noop {} {\bibfield
  {journal} {\bibinfo  {journal} {Comput. Phys. Commun.}\ }\textbf {\bibinfo
  {volume} {224}},\ \bibinfo {pages} {81} (\bibinfo {year} {2018})}\BibitemShut
  {NoStop}%
\end{thebibliography}%
\clearpage

\end{document}